\pdfoutput=1
\documentclass[twocolumn]{aastex61}

\received{}
\revised{}
\accepted{}
\submitjournal{ApJ}

\shorttitle{ISM in protocluster at z $\sim$ 2.5}
\shortauthors{M. M. Lee et al.}
\begin{document}
\title{A radio-to-mm census of star-forming galaxies in protocluster 4C23.56 at z=2.5 :\\ Gas mass and its fraction revealed with ALMA}

\correspondingauthor{Minju M. Lee}
\email{minju.lee@nao.ac.jp}

\author[0000-0002-2419-3068]{Minju M. Lee}
\affiliation{Department of Astronomy, The University of Tokyo, 7-3-1 Hongo, Bunkyo-ku, Tokyo 133-0033, Japan}
\affiliation{National Observatory of Japan, 2-21-1 Osawa, Mitaka, Tokyo 181-0015, Japan}

\author{Ichi Tanaka}
\affiliation{Subaru Telescope, National Astronomical Observatory of Japan, 650 North Aohoku Place, Hilo, HI 96720, USA}

\author{Ryohei Kawabe}
\affiliation{Department of Astronomy, The University of Tokyo, 7-3-1 Hongo, Bunkyo-ku, Tokyo 133-0033, Japan}
\affiliation{National Observatory of Japan, 2-21-1 Osawa, Mitaka, Tokyo 181-0015, Japan}
\affiliation{SOKENDAI (The Graduate University for Advanced Studies), 2-21-1 Osawa, Mitaka, Tokyo 181-0015, Japan}

\author{Kotaro Kohno}
\affiliation{Institute of Astronomy, The University of Tokyo, 2-21-1 Osawa, Mitaka, Tokyo, 181-0015, Japan}
\affiliation{Research Center for the Early Universe, The University of Tokyo, 7-3-1 Hongo, Bunkyo, Tokyo 113-0033}

\author{Tadayuki Kodama}
\affiliation{Astronomical Institute, Tohoku University, Aoba-ku, Sendai 980-8578, Japan}

\author{Masaru Kajisawa}
\affiliation{Graduate School of Science and Engineering, Ehime University, Bunkyo-cho, Matsuyama 790-8577, Japan}
\affiliation{Research Center for Space and Cosmic Evolution, Ehime University, Bunkyo-cho, Matsuyama 790-8577, Japan}

\author{Min S. Yun}
\affiliation{Department of Astronomy, University of Massachusetts, Amherst, MA 01003, USA}

\author{Kouichiro Nakanishi}
\affiliation{National Observatory of Japan, 2-21-1 Osawa, Mitaka, Tokyo 181-0015, Japan}
\affiliation{SOKENDAI (The Graduate University for Advanced Studies), 2-21-1 Osawa, Mitaka, Tokyo 181-0015, Japan}

\author{Daisuke Iono}
\affiliation{National Observatory of Japan, 2-21-1 Osawa, Mitaka, Tokyo 181-0015, Japan}
\affiliation{SOKENDAI (The Graduate University for Advanced Studies), 2-21-1 Osawa, Mitaka, Tokyo 181-0015, Japan}

\author{Yoichi Tamura}
\affiliation{Department of Physics, Nagoya University, Furo-cho, Chikusa-ku, Nagoya 464-8601, Japan}

\author{Bunyo Hatsukade}
\affiliation{Institute of Astronomy, The University of Tokyo, 2-21-1 Osawa, Mitaka, Tokyo, 181-0015, Japan}

\author{Hideki Umehata}
\affiliation{Institute of Astronomy, The University of Tokyo, 2-21-1 Osawa, Mitaka, Tokyo, 181-0015, Japan}
\affiliation{The Open University of Japan, 2-11 Wakaba, Mihama-ku, Chiba 261-8586, Japan}

\author{Toshiki Saito}
\affiliation{National Observatory of Japan, 2-21-1 Osawa, Mitaka, Tokyo 181-0015, Japan}

\author{Takuma Izumi}
\affiliation{National Observatory of Japan, 2-21-1 Osawa, Mitaka, Tokyo 181-0015, Japan}

\author{Itziar Aretxaga}
\affiliation{Instituto Nacional de Astrofisica, Optica y Electronica (INAOE), Aptdo. Postal 51 y 216, 72000 Puebla, Mexico}

\author{Ken-ichi Tadaki}
\affiliation{National Observatory of Japan, 2-21-1 Osawa, Mitaka, Tokyo 181-0015, Japan}

\author{Milagros Zeballos}
\affiliation{Instituto Nacional de Astrofisica, Optica y Electronica (INAOE), Aptdo. Postal 51 y 216, 72000 Puebla, Mexico}
\affiliation{Instituto Tecnologico Superior de Tlaxco, Predio Cristo Rey Ex-Hda de Xalostoc s/n, 90250 Tlaxcala, Mexico}

\author{Soh Ikarashi}
\affiliation{Kapteyn Astronomical Institute, University of Groningen, P.O. Box 800, 9700 AV Groningen, The Netherlands}

\author{Grant W. Wilson}
\affiliation{Department of Astronomy, University of Massachusetts, Amherst, MA 01003, USA}

\author{David H. Hughes}
\affiliation{Instituto Nacional de Astrofisica, Optica y Electronica (INAOE), Aptdo. Postal 51 y 216, 72000 Puebla, Mexico}

\author{R. J. Ivison}
\affiliation{Institute for Astronomy, University of Edinburgh, Royal Observatory, Blackford Hill, Edinburgh EH9 3HJ, UK}
\affiliation{European Southern Observatory, Karl-Schwarzschild-Str. 2, D-85748 Garching, Germany}



\begin{abstract}
We investigate gas contents of star-forming galaxies associated with protocluster 4C23.56 at z = 2.49 by using the redshifted CO~(3--2) and 1.1 mm dust continuum with the Atacama Large Millimeter/submillimeter Array.
The observations unveil seven CO detections out of 22 targeted H$\alpha$ emitters (HAEs) and four out of 19 in 1.1 mm dust continuum. 
They have high stellar mass ($M_{\star}>4\times 10^{10}$ $M_{\odot}$) and exhibit a specific star-formation rate typical of main-sequence star forming galaxies at $z\sim2.5$. 
Different gas mass estimators from CO~(3--2) and 1.1 mm {yield} consistent values for simultaneous detections.
The gas mass ($M_{\rm gas}$) and gas fraction ($f_{\rm gas}$) are comparable to {those of} field galaxies, {with} $M_{\rm gas}=[0.3, 1.8]\times10^{11} \times (\alpha_{\rm CO}/(4.36\times A(Z)$)) M$_{\odot}$, where $\alpha_{\rm CO}$ is the CO-to-H$_2$ conversion factor and $A(Z)$ the additional correction factor for the metallicity dependence of $\alpha_{\rm CO}$, and $\langle f_{\rm gas}\rangle = 0.53 \pm 0.07$ from CO~(3--2).
Our measurements place a constraint on the cosmic gas density of high-$z$ protoclusters, indicating the protocluster is characterized by a gas density higher than that of the general fields by an order of magnitude. We found $\rho (H_2)\sim 5 \times 10^9 \,M_{\odot}\,{\rm Mpc^{-3}}$ with the CO(3-2) detections.
The five ALMA CO detections occur in the region of highest galaxy surface density, where the density positively correlates with global {star-forming} efficiency (SFE) and stellar mass.
Such correlations imply a potentially critical role of environment on early galaxy evolution at high-z protoclusters, although future observations are necessary for confirmation.
\end{abstract}
\section{Introduction} \label{sec:intro}
In the last four decades, it has become clear that galaxy evolution is intertwined with the surrounding environment. 
Galaxy properties such as star-formation rate, color, and morphology are strongly correlated with projected number densities
(e.g., \citealt{Dressler1980, Dressler1997, Balogh1998, Baldry2004, Kauffmann2004, Blanton2005, Poggianti2008, Vulcani2010, Wetzel2012}, see also \citealt{Blanton2009} for a review). 
It is also acknowledged that the fraction of blue star-forming galaxies increases in clusters with increasing redshift (so-called Butcher-Oemler effect, \citealt{Butcher1978, Butcher1984}).
These observations are the results of the gas supply that fuels the galaxy, and its consumption or removal (e.g., via feedback and/or stripping). 
These are functions of the environment (defined by galaxy number density or the distance to the 5th member, for example, to trace dark matter halo) where {complex} hydrodynamical mechanisms of baryons and gravitational forces of dark matter are working behind.

Typical star-forming galaxies are generally defined on the plane of SFR-M$_{\star}$, and the normalization factor, the specific star formation rate (sSFR) of such star-forming galaxies, evolves as a function of redshift  (e.g., \citealt{Noeske2007, Daddi2007, Whitaker2012, Speagle2014, Kurczynski2016}) and increases with redshift at least up to $z\sim6$ with fairly tight scatter ($\sim$0.3 dex).
Therefore, more stars are formed in galaxies at higher redshift {and }at {a} given stellar mass. 
With the advent of large surveys revealing the gas content of star forming galaxies, the evolution of sSFR appears to be caused by the higher gas fraction ($f_{\rm gas}=M_{\rm gas}/(M_{\rm gas}+M_{\star}$)), rather than a higher efficiency {of transformation of }gas into a star, at least on the main sequence. 
{Further,} the higher $M_{\rm gas}$ appears to mimic the higher gas supply rate (e.g., \citealt{Tacconi2013, Magdis2012, Saintonge2013, Sargent2014}; \citealt{Genzel2015}, hereafter G15; \citealt{Scoville2014, Scoville2016}, hereafter S16; \citealt{Schinnerer2016})

Since galaxies evolve not only as a function of redshift but also of their environment,
one needs to understand how the gas content and its fraction changes with {the} environment, from fields to groups to clusters across the cosmic time (where sSFR also evolves).
With such understanding, we can determine whether star-forming processes are different or similar, e.g., in terms of global star-forming efficiency (SFE) or depletion time scale $\tau_{\rm depl.}$=1/SFE.
This allows us to understand {the} physical mechanism driving galaxy evolution in different environments.
{Information on the} gas content and its fraction {is insufficient from} environmental perspectives, specifically for high redshift ($z\gtrsim2$) clusters and their ancestors, i.e., protoclusters.
At z=0, there is a large number of not only HI but also CO gas (to probe $H_2$) surveys (e.g., \citealt{Boselli2014, Cybulski2016}).
Although the number of observations of clusters, groups, and voids is increasing, it is still limited, we are now beginning to understand how gas content changes as a function of environment at {a} fixed redshift (e.g., \citealt{Chamaraux1980, Leon1998, Cortese2008, Chung2009, Serra2012, Boselli2014, Das2015, Alatalo2015, Mok2016}).

Direct measurements of gas content {of} high-$z$ (proto)cluster members are still limited to one or two samples of starbursts\footnote{We hereafter {use the term "starburst" to refer to} a galaxy well above the main sequence ($>0.6$ dex), {which} may include {\it classical} submillimeter bright (i.e., $S_{850~\mu \rm m}>5 $ mJy) galaxies (SMGs).
To be clear, SMGs {refer to galaxies} generally detected with a submm single dish previously, {which are} thus unresolved and {are} a subpopulation of starbursts within {this} paper.
We explicitly {use the term} "main-sequence SMGs" when these galaxies are, once resolved, on the main sequence (with smaller flux densities) . 
This is to follow recent higher-resolution follow-up observations with ALMA {demonstrating} that such classical SMGs are divided into subgroups of starbursts and the main sequence when they are resolved (\citealt{da Cunha2015}).}
(i.e., well above the main sequence $>0.6$ dex) (e.g., \citealt{Riechers2010, Tadaki2014}) {and }AGNs (e.g., \citealt{Emonts2013}) per system. 
These rare populations are known to be more abundant in high-z overdensities than in general fields at the same redshift (e.g., \citealt{Lehmer2013, Umehata2015, Casey2016} and references therein).
They are relatively easy to detect given their extreme nature (i.e., high SFR, brightness and/or richness of dust (submm-bright)).
While the existence of these {populations} within high-z overdensities may play a profound role in galaxy evolution during cluster-formation epoch, it is necessary to constrain the properties of typical (i.e., on the main sequence) star-forming galaxies to fully construct the picture of galaxy evolution, since they are a dominant population.
There is no significant direct detection of molecular lines or dust continuum of the main-sequence galaxies in protoclusters (e.g., \citealt{Hodge2013}, but see \citealt{Chapman2015} (and references therein) for a report of the detection {of a} normal (UV-faint) galaxy on the main sequence, with possibly CO~(3--2) line emission), even though main-sequence star forming galaxies have been reported to be dustier in {a} high-$z$ (proto)cluster (e.g., \citealt{Koyama2013}). 

In this paper, we reveal for the first time the gas content and its fraction of star forming galaxies that are securely associated to a protocluster at z=2.49, where multi-band ancillary data sets are available, as a case study.
The term gas hereafter refers to the molecular gas as the measurement is, but {it} can {be regarded} {\it effectively} {as} the total gas mass at the considered resolution ($\sim$a few kpc) {because} the atomic gas content might be negligible (within the effective radius) with higher ISM pressure at high redshift, particularly {at} the massive-end (e.g., \citealt{Obreschkow2009a, Lagos2012}).

This {is} the first paper in a series of papers that will unveil the properties of star-forming galaxies associated to the protocluster 4C23.56 at z=2.49.
In this Paper, we directly observe both gas, i.e., CO~(3--2), and dust, meaning that we derive gas content {\it without} using {a} SFR-based empirical relation {such as the} Kennicutt--Schmidt (KS)-relation (e.g., \citealt{Schmidt1959, Kennicutt1998b}). 
This allows us to overcome uncertainties included in the conversion from SFR to gas mass and to check the consistency between two different measurements.
Currently scheduled subsequent papers {will report} i) the kinematics and structural properties of the galaxies combined with higher-resolution imaging (M. Lee et al., in preparation) and ii) UV-to-radio SED fitting and AGN contribution by adding X-ray (Chandra), mid-infrared (from IRAC and MIPS), and radio (Jansky Very Large Array; JVLA) data sets (M. Lee et al., in preparation).

The remainder of this Paper is {organized} as follows. 
We illustrate the sample selection and introduce our target field in Section~\ref{sec:sample}.
In Section~\ref{sec:observation}, we present details of the observations, data reduction, imaging and analysis of the ALMA data. 
Section~\ref{sec:multiband} {presents} a brief summary of ancillary data sets that are discussed within the Paper.
In Section~\ref{sec:barmass}, we present the measurements of barynoic gas mass and its fraction.
We finally discuss the results by focusing on the different and similar properties found in the protocluster star-forming galaxies in Section~\ref{sec:discussions}.
A summary is given in Section~\ref{sec:summary}.

Throughout this paper, we assume $H_0 = 67.8$ ${\rm km\,s^{-1}\,Mpc^{-1}}$, $\Omega_0 = 0.308$ and $\Omega_{\Lambda} = 0.692$ \citep{Planck Collaboration2015}.
The adopted initial mass fuction (IMF) is Chabrier IMF in the mass range 0.1 - 100 M$_{\odot}$.

\section{Sample selection and target field}\label{sec:sample}
\subsection{H$\alpha$ emitters}
We targeted H$\alpha$ emitters (HAEs) that were originally detected using the narrow band (NB) technique (\citealt{Tanaka2011}, I. Tanaka et al., in preparation) with MOIRCS/Subaru (\citealt{Ichikawa2006}). 
In the parent sample, 25 HAEs {were} detected within the field of view (FoV) ($\sim28$ arcmin$^2$, corresponding to $\sim 84$ co-moving Mpc$^2$) of MOIRCS/Subaru.
They are most likely associated to protocluster 4C23.56, given the width of the NB filter{, which is} $\Delta \lambda$ = 0.023~$\mu$m with {a} central wavelength of 2.288~$\mu$m so that the H$\alpha$ emission {can be traced} within $2.469 < z < 2.503$ ($\sim$40 comoving Mpc).
The redshift range corresponds to the velocity width {of} $\pm 1500$~km~s$^{-1}$, which is {sufficiently} large to trace the non-virialized protocluster members. 
For reference, the velocity dispersion of Lyman alpha emitters (LAEs) associated to protoclusters at $z=2-3$ is $\sim200-1000$~km~s$^{-1}$ (\citealt{Venemans2007, Chiang2015}). 
From simulations, the expected size of high-z {protoclusters} near $z=2-3$ is $R_e\sim 5-10$ comoving Mpc depending on the size at $z=0$ (\citealt{Chiang2013, Muldrew2015}).

\begin{figure}[tb]
\includegraphics[width=0.5\textwidth, bb = 0 0 1024 789]{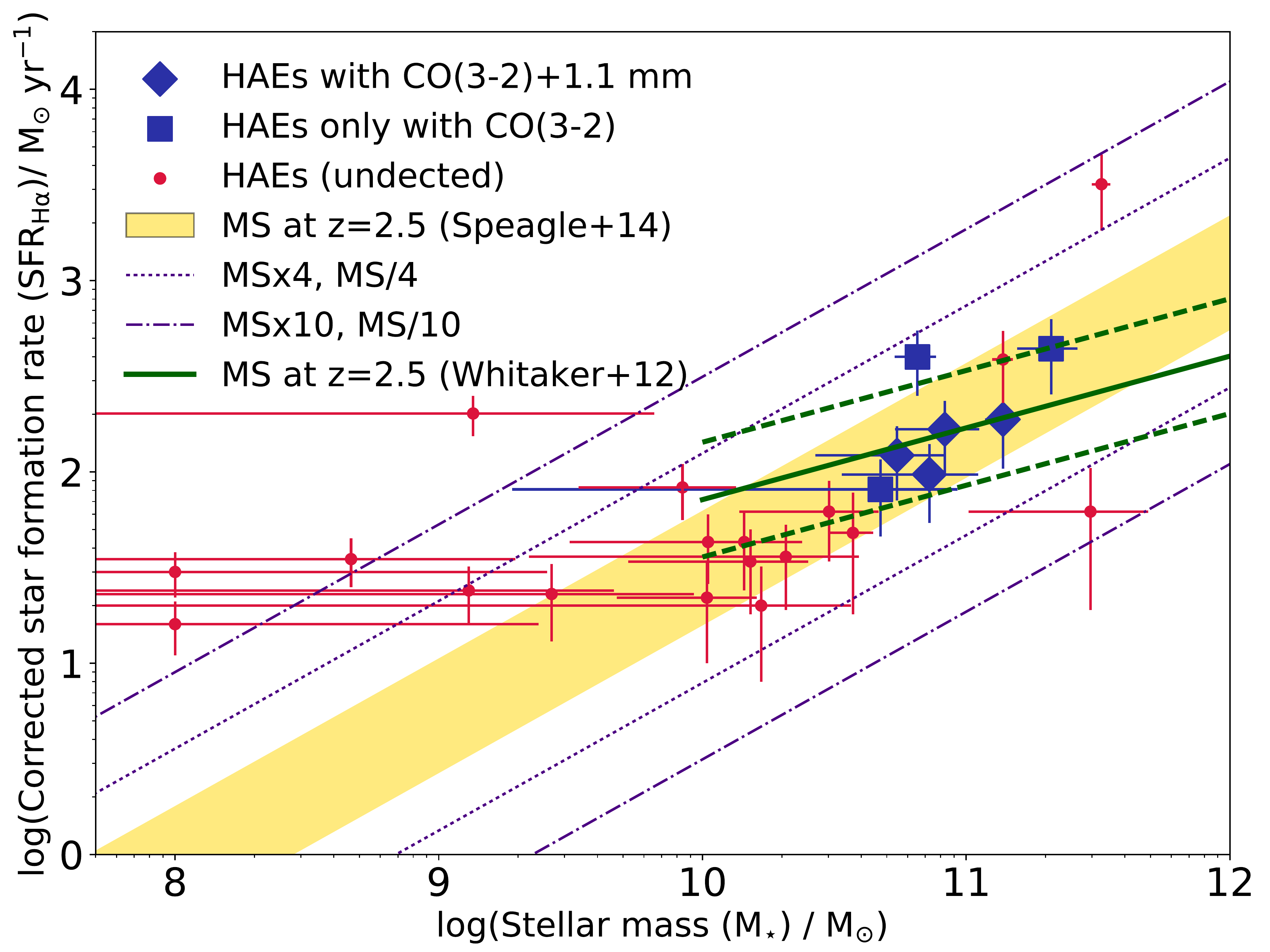}
\caption{Distribution of galaxies in the SFR--$M_{\star}$ plane of the parent samples of HAEs (I. Tanaka et al., in preparation). The stellar mass is derived from {the} J {and} Ks bands and SFR is derived from the (continuum subtracted) NB flux by considering dust extinction and [NII] contribution (see also Sec~\ref{sec:subaru} for a short description). We also plot lines for galaxies above ($\times4, \times10$, dotted) and below (1/4, 1/10, dashed dot) the main sequence at z=2.5. We used formulae presented in \citet{Speagle2014} (yellow band) and \citet{Whitaker2012} (green solid line and dashed lines for $\pm0.3$ dex) to show the z$\sim$2.5 main sequence galaxies. Most HAEs with stellar mass of M$\star > 10^{10}$ M$_{\odot}$ are on the main sequence within the scatter of the main-sequence galaxies ($\pm0.3$ dex), which will be the main targets discussed in this paper. \label{fig:galaxysequence}}
\end{figure}

The HAEs in the parent sample spans three orders of magnitude in $M_{\star}$ and two orders of magnitude in SFR (Fig.~\ref{fig:galaxysequence}, I. Tanaka et al., in preparation) ($0.2<{\rm sSFR (Gyr^{-1}})<301.0$, {and} the typical sSFR of the main sequence is $\sim1-3$ (Gyr$^{-1}$) (e.g., \citealt{Whitaker2012, Speagle2014}). 
In particular, the massive ($\gtrsim 10^{10}$ $M_{\odot}$) galaxies mainly discussed in this paper are mostly on the main sequence.
 As such, the (NB-selected) HAEs have been studied to investigate the nature of typical (massive) star-forming galaxies on the main sequence (e.g., \citealt{Geach2008, Sobral2009, Koyama2013, Tadaki2013, Oteo2015}).

\begin{figure*}
\includegraphics[width = 17 cm, bb = 0 0 1024 789]{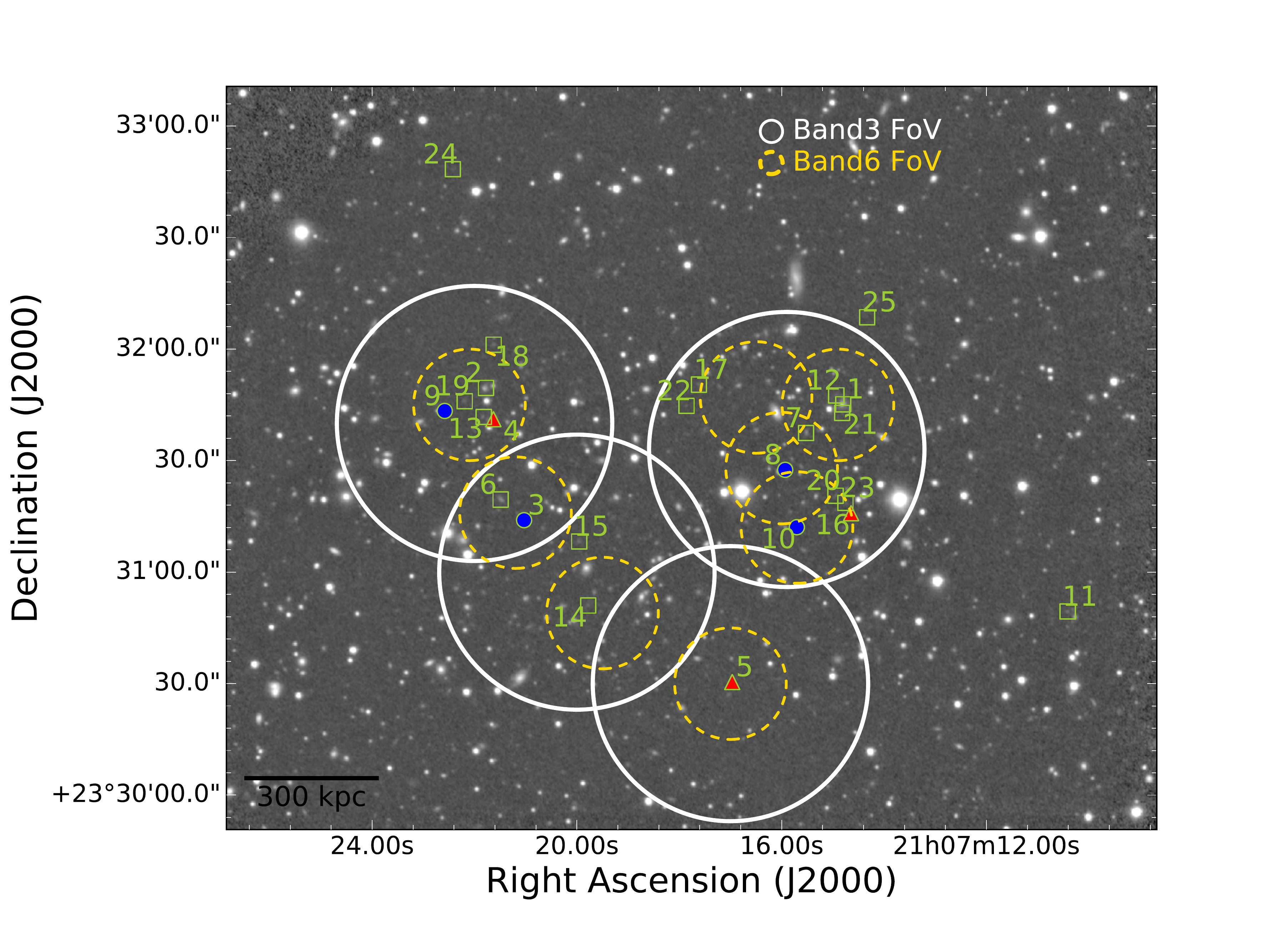}
\caption{The distribution of HAEs tagged by the source ID, overlaid on the Subaru/MOIRCS Ks band image (I. Tanaka et al., in preparation).
The blue filled circles indicate galaxies detected simulataneously in CO~(3--2) and 1.1 mm, red triangles indicate galaxies only in CO~(3--2). Green open squares show the remainder of HAEs detected with the NB filter technique.
The fields of view (FoVs) of ALMA Band~3 CO~(3--2) (white open circles) and Band~6 1.1 mm (yellow dashed circles) observations are shown on the map. 
The total number of pointing is 4 and 8 for Band~3 and Band~6, respectively. 
A scale bar is shown at the bottom left corner to represent a physical size of 300 kpc.\label{fig:fovs}
}
\end{figure*}

We observed the HAEs with Atacama Large Millimeter/submillimeter Array (ALMA). 
The Band~3 CO~(3--2) observations have been performed to cover 22 HAEs, and the Band~6 1.1 mm observations {have been performed} to cover 19 HAEs (See Fig.~\ref{fig:fovs}). 
The targeted and detected numbers, while limited to the field coverage, constitute the largest sample of typical star-forming galaxies on the the main sequence associated to the protocluster that are probed for emission-line and dust-continuum observations.
We have listed ALMA-targeted samples in Table~\ref{tab:physicparam} and~\ref{tab:nondet}. 
The IDs in the first column are revised versions of those in \citet{Tanaka2011}, and the reference IDs from \citet{Tanaka2011} {are shown} in the last column.
%%%%%%%%%%%%%%%%%%%%
%%%%FIGURES 1,2 starts%%%%%%

%%%%%%%%%%%%%%%
%
\subsection{Protocluster 4C23.56}
Protocluster \objectname{4C23.56} was identified as an overdense region of the NB-selected HAEs that was a part of {the} MAHALO-Subaru (MApping HAlpha and Lines of Oxygen with Subaru) survey \citep{Kodama2015}. 
Radio galaxy \objectname{4C23.56} (HAE1) at z = 2.483 $\pm$ 0.003 is associated to this protocluster (\citealt{Roettgering1997}).
Historically, radio galaxies have been targeted in a search for (proto)clusters since their hosts are the most massive galaxies \citep{Seymour2007} and {are} expected to be embedded in the most massive halos (e.g., \citealt{Rocca-Volmerange2004, Orsi2016}). 
Indeed, the method has successfully yielded promising results to find (proto)clusters (e.g., \citealt{Le Fevre1996, Kurk2000, Best2003, De Breuck2004, Overzier2006, Venemans2007, Hatch2011}) and protocluster \objectname{4C23.56} is also one of them.

Protocluster \objectname{4C23.56} is known to have overdensities of differently selected galaxy populations besides HAEs \citep{Tanaka2011}. 
In other words, the protocluster is rich in ancillary data that ranges from X-ray to radio ; therefore, {it} is one of the best targets to study the properties of typical star-forming galaxies in protocluster regions.
Currently, the protocluster has been known to have (projected) overdensities of, for example, mass-selected distant red galaxies (DRGs) \citep{Kajisawa2006}, extremely red objects (EROs) \citep{Knopp1997}, IRAC \citep{Mayo2012}, MIPS \citep{Galametz2012} sources and SMGs observed at 1.1 mm with the Atacama Submillimeter Experiment (ASTE; K. Suzuki 2013 PhD thesis; M. Zeballos et al. in preparation). 
These populations, however, have only rough (e.g., lower limit) or no redshift constraints compared to the relatively secure narrow redshift range of HAEs from the NB technique.

\begin{figure*}
\includegraphics[width = 17 cm, bb = 0 0 1024 789]{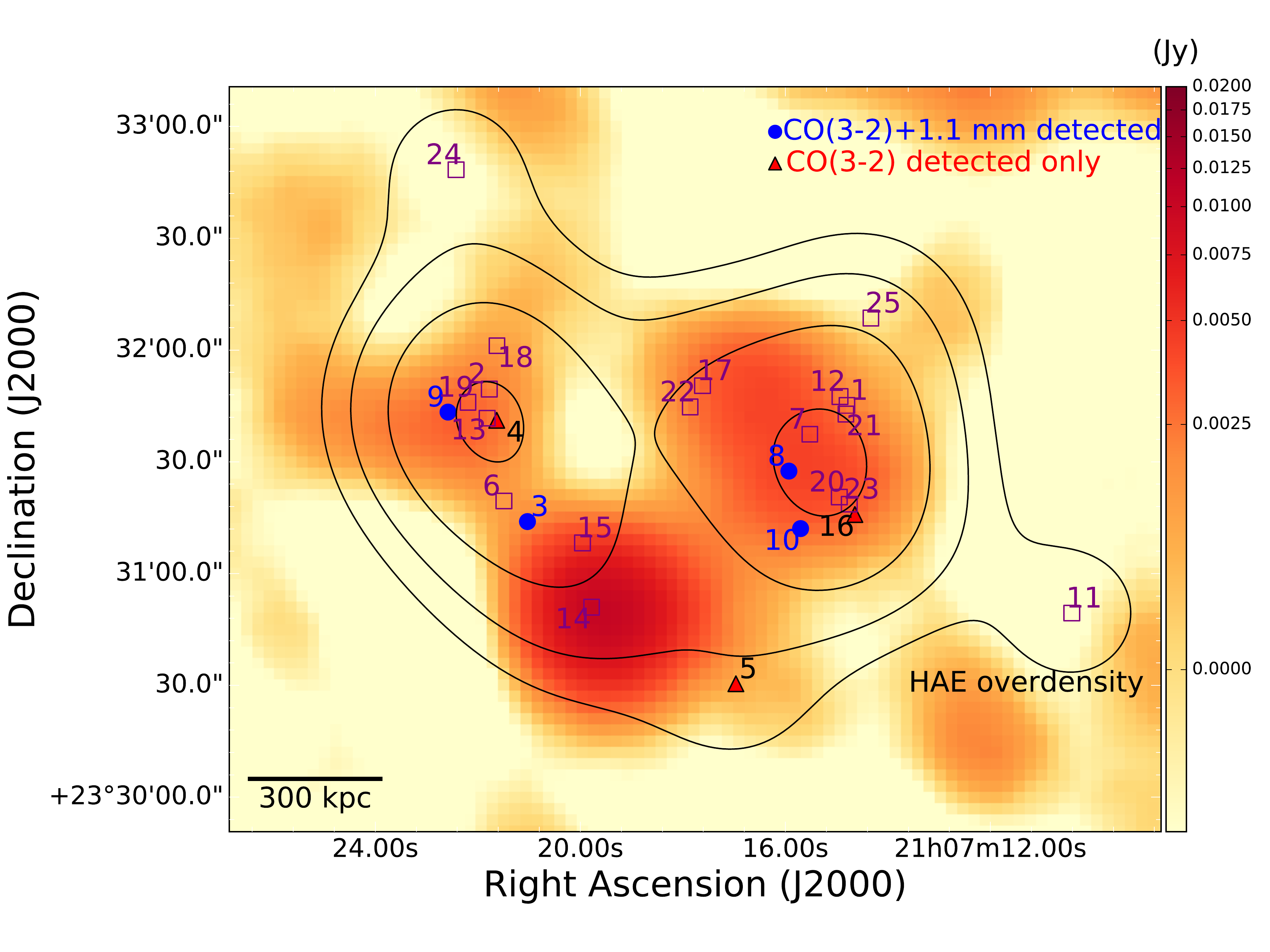}
\caption{The distribution of HAEs tagged by the source ID, overlaid on the AzTEC/ASTE 1.1 mm single dish image (background color, K. Suzuki 2013 PhD thesis; M. Zeballos et al. in preparation). 
Multiple SMGs are nicely overlapped with HAEs suggesting that HAEs are undergoing a dusty star formation.
The brightest SMG (4C23-AzTEC 1) detected with AzTEC/ASTE at 1.1 mm, near HAE14, is not associated to the protocluster (Suzuki et al., in preparation). 
Thus, relatively moderate star forming galaxies on the main sequence appear to be associated to the protocluster. 
Four blue filled circles are for galaxies detected simultaneously in CO~(3--2) and 1.1 mm, red triangles for CO~(3--2) only detection, and purple squares for the rest of the HAEs.
The ALMA observations have confirmed the association of 1.1 mm dust continuum emission for four  HAEs (HAE3, 8, 9, 10).
The number next to the color bar on the right is written in the unit of Jy to show the flux level of the 1.1 mm AzTEC sources. 
We also plot a HAE surface overdensity map in black contours that is estimated by assuming a Gaussian kernel with a radius of $0^{\prime}.8$ that corresponds to the physical size of 400 kpc in radius (or $\sim$1.4 comoving Mpc in radius), in steps of [1, 2, 4, 8] (arbitrary unit)(Sec~\ref{sec:dense}). 
\label{fig:aste}
}
\end{figure*}

Nonetheless, some populations have several indirect evidences that imply association with the protocluster.
For example, three SMGs discovered with ASTE overlap with the position of all of our HAEs except for HAE5, 11, 24 and 25 (Fig.~\ref{fig:aste}).
The positions of three SMGs are also roughly coincident with the peak overdensity of the HAEs (with a resolution of $\sim30^{\prime\prime}$). 
This has prompted an idea that HAEs associated to the protocluster are experiencing a dusty star forming phase and the SMGs are associated to the protocluster. 
We followed up the HAEs (and the SMGs with overlaps) with ALMA, which allows to pin down {the} 1.1 mm continuum.
%%%Figure2 starts%%%%
%%%%%%%%%%%%%%%%%%%%%
%%%%%%%%%%%%%%%
%%%%Figure3 starts%%%%%%

\section{ALMA observations and analysis}\label{sec:observation}
\subsection{CO~(3--2) at Band~3 and 1.1 mm at Band~6}
ALMA 1.1 mm observations were performed in Cycle 1 and CO~(3--2) {observations were performed} in Cycle 2 (ALMA\#2012.1.00242.S, PI: K. Suzuki).

The Band~6 continuum observations at 1.1 mm were conducted with a total on-source time of $\sim30$ mins for 8-pointing target observations (typically $\sim4$ mins per pointing direction), covering 19 out of 25 HAEs (Fig.~\ref{fig:fovs}).
The correlator is set to target four spectral windows with {an} effective bandwidth of $\sim1.875$ GHz each that is taken in {the} time division mode (TDM) (channel widths of 15.6 MHz or $\sim18$ km s$^{-1}$). 
The central frequencies of the four spectral windows are 256.0, 258.0, 272.0, and 273.8 GHz.
The noise level (1$\sigma$) reached $\sim0.08~$mJy~beam$^{-1}$ per field, except {for} one {case} with $\sim0.12$~mJy~beam$^{-1}$ where a bright SMG 4C23 AzTEC1 SMG ($S_{\rm1.1\, mm,\,single\,dish}=10$ mJy) is located (K. Suzuki 2013 PhD thesis; M. Zeballos et al. in preparation). 
The baseline lengths were between 17 and 462~m. 
We observed J2148+0657, Neptune, and J2025+3343 as a bandpass, flux, and phase calibrator, respectively.

The Band~3 CO(3-2) observations were executed for {a} total {of} 4 hours of on-source time with 4-pointing (thus $\sim1$~hr per pointing {direction}), targeting 22 HAEs. 
The correlator is set to target four spectral windows with effective bandwidth of $\sim1.875$~GHz each. One of the spectral windows, centered at 99.3 GHz, is taken in frequency division mode (FDM) (channel widths of 0.49 MHz or $\sim 1.5$~km~s$^{-1}$), where the redshifted CO(3--2) line ($\nu_{\rm rest}$ = 345.79599 GHz) at $z=2.5$ would fall, while the remaining three spectral windows are taken in {the} TDM mode (a channel width of 15.6~MHz or 47~km~s$^{-1}$) and {are} centered at 101.1, 111.3 and 113.2 GHz.
The velocity coverage of the CO observations is $\sim6400$~km~s$^{-1}$, corresponding to a redshift coverage of $2.385 < z < 2.516$ in {the} lower side band and $2.031 < z < 2.134 $ in {the} upper side band. 
This is sufficient to cover the expected redshift range of the 22 HAEs detected by the NB technique ($z=2.486 \pm 0.017$).
All 19 HAEs covered by the 1.1 mm observations were fully covered by the Band~3 observations (Fig.~\ref{fig:fovs}). 
The typical noise (1$\sigma$) level reached $\sim 0.17$~mJy when the spectral resolution is re-binned to 100~km~s$^{-1}$. 
We chose a spectral resolution of 100~km~s$^{-1}$ to estimate the signal-to-noise ratio (S/N) (as {the} detection criteria, {see} Sec~\ref{sec:detectcriteria}) and upper limits for non-detection, except {for} a case in which we needed {a} higher velocity resolution.
For example, the treatment was applied for HAE5 since we found strong emission in a single channel with S/N $>$ 6.5.
Thus, we re-imaged the source with a spectral resolution of 30~km~s$^{-1}$ and found that the fitted line width is FWHM $\sim 100$~km~s$^{-1}$. 
The noise level in this case became worse, i.e., $\sim0.3$~mJy, but it was sufficient in that the detection of this galaxy satisfied our detection criteria (see Section~\ref{sec:detectcriteria}).
The flux calibrator was Titan for Band~3. 
J1751+0939 and J2148+0657 were chosen as bandpass calibrators and 
J2025+3343 as a phase calibrator.
The minimum baseline was 43~m, and the maximum baseline {was} 1574~m for Band~3.

We applied {the} CLEAN algorithm to the calibrated visibilities {with natural weighting} to produce images for both observations {by} using {the} Common Astronomy Software Applications package (CASA, used 4.2.2 version for calibration and imaged with 4.6.0 version).
The absolute flux uncertainties for both bands were estimated {as}$ \sim15-18$\%, which were not taken into account for the flux error throughout this paper.
The synthesized beam sizes are
0$^{\prime\prime}$.91$\times$0$^{\prime\prime}$.66 (PA = $23.5^{\circ})$ for Band~3 and
0$^{\prime\prime}$.78$\times$0$^{\prime\prime}$.68 (PA = $0.4^{\circ}$) for Band~6.
The sub-arcsec resolution is sufficient to pin down SMGs detected by ASTE (with its typical beam size of $\sim 30^{\prime\prime}$)
and to search for counterparts detected at other wavelengths, e.g., images obtained in NIR/optical bands. 

\subsection{Detection and flux measurement}
\subsubsection{Detection criteria}\label{sec:detectcriteria}
We searched for emissions around the position of HAEs with a searching radius of $r=1^{\prime\prime}$. 
We regarded a galaxy as detected in ALMA Band~6 (1.1 mm continuum) if a peak flux density is above $4~\sigma$.
A CO~(3--2) line was regarded as detected if at least two among three criteria (a-c) are satisfied : 
(a) a peak flux $>4~\sigma$, 
(b) at least two continuous channels including a maximum peak flux channel have {flux} $> 3.5~\sigma$, 
{and }(c) (spatially smoothed) velocity-integrated peak flux S/N is above 5 before the primary beam correction. 
All galaxies except HAE4 (=6/7) satisfy {all} the conditions. 
HAE4 has {two} distinct {but} not continuous peaks ($> 4~\sigma$) {that are} 100 km s$^{-1}$ (one channel) apart (Fig.~\ref{fig:postage1}).
We show CO~(3--2) spectra in Fig.~\ref{fig:postage1}-\ref{fig:postage4}, but a detailed analysis that deals with the kinematics and sizes is beyond the scope of this paper and will be presented in a subsequent paper (M. Lee et al., in preparation). 

We note that the detection is not a false identification of spurious or other lines at {a} different redshift, provided the redshift range of {the} NB filter and our on-going parallel NIR spectroscopy using the upgraded MOIRCS (`nuMOIRCS') aboard Subaru. 
The spectroscopic campaign has {thus far} confirmed the redshifts of 15 HAEs that are all within $z=2.49\pm0.01$ (I. Tanaka et al., in preparation).
We defined the CO~(3--2) redshift from the median velocity component due to the broad nature of spectrum for many of the galaxies.
The CO redshift is consistent with the NIR spec-$z$ value within an error of $\Delta z = 0.004$ for most of the cases, but $\Delta z = 0.01$ for HAE3 and HAE4 owing to {the} low S/N in the NIR spectroscopy.
\subsubsection{Flux}\label{sec:flux}
We adopted a peak flux at 1.1 mm or a peak velocity-integrated flux in {the} CO~(3--2) moment 0 map to compute gas mass, which was measured from a smoothed map.
We measured the flux after primary beam correction.
All of the sources are within a good sensitivity region{ ;} thus, the measured flux was not changed significantly by the primary beam correction (within $\sim10\%$). 
The noise level for the flux uncertainty in Band~3 was estimated by averaging five line-free channels in the primary-beam-corrected-image, which was cut out around the source (with a size of $15\times15^{\prime\prime}$) using {\texttt{immath}} from the original map with FoV of $\sim74^{\prime\prime}$ and then masked (with a radius of 1.5 times the beam size) for the known bright sources (including HAEs). 
Similarly, the noise level in Band~6 was derived from the image {sliced} around the source with a size of $6^{\prime\prime}\times6^{\prime\prime}$ from a larger image with FoV $\sim30^{\prime\prime}$ and then masked for the detected known sources.

We smoothed images using {the} CASA command {\texttt{imsmooth}}. 
This treatment was performed to neglect a galaxy structure for the measurement of global gas content. 
We found that image-based smoothing delivers a better S/N than tapering the uv visibilities.
In addition, smoothing allows us to avoid the divergence of 1-component Gaussian spectrum fitting for a disturbed galaxy, which likely constitute roughly a half of the detected HAEs. 
The images in Band~3 for CO~(3--2) were smoothed channel by channel.
By making a measurement from the smoothed map, we could also maximize the S/N by collecting diffuse, extended emissions from the outskirts of a galaxy that could be missed with the sub-arcsec ($\sim$ 6 kpc at $z=2.5$) beam size. 

We adopted a smoothing Gaussian kernel {size} of $0^{\prime\prime}.8\times0^{\prime\prime}.8$ for Band~6 and $0^{\prime\prime}.6\times0^{\prime\prime}.6$ for Band~3.
A detailed analysis of the choice of Gaussian kernels is presented in Appendix~\ref{app:flux}.
In brief, we investigated S/N as a function of the smoothing kernel, which is effectively equivalent to considering the growth curve of galaxy emission as a function of aperture size.
This results in a similar smoothed beam size of $1^{\prime\prime}.1\times0^{\prime\prime}.9$ for Band~3 and 
$1^{\prime\prime}.1\times1^{\prime\prime}.0$ for Band~6.
We find that, at the adopted beam sizes, the S/N is maximum and the flux is $\sim 50-90\%$ of the maximum flux measured up to $4^{\prime\prime}.0$ ({physical size of }$\sim$ 33 kpc at $z=2.5$) smoothing kernel.
We show the growth curves as a function of {the} smoothing Gaussain kernel in Fig.~\ref{fig:growthcurveb3} and \ref{fig:growthcurveb6} to show {that} the adopted kernel is not a bad choice. 
We note that some galaxies have a low recovery flux with respect to the maximum peak value, but {all} these have a relatively low S/N ; therefore, the uncertainty is also large in the absolute flux. 
Thus, we opt to choose the universal smoothing kernels for the analysis.
The beam sizes correspond {to} $\sim8.5$ kpc in physical scale for both 1.1 mm and CO~(3--2) and are sufficient to recover the total flux given the typical size of {a} star-forming galaxy at high $z$ ($r_{1/2,\rm CO}$ $\sim$ 5 kpc, e.g., \citealt{Bolatto2015}).

Instead of performing a Gaussian fit for CO~(3--2), we compute and choose an integrating range for CO~(3--2) to obtained the maximum S/N in the peak flux in the velocity integrated image following the description in \citet{Seko2016}. 
The map was checked by eye afterward for unexpected cases, such as extremely broad line widths to integrate, because some galaxies have unusual spectra that are not well-fitted with a single gaussian, in addition to inhomogeneous spatial distributions and the velocity gradients ({see }Fig.~\ref{fig:postage1}-\ref{fig:postage4} for the morphology and spectrum, M. Lee et al., in preparation). 

With our detection criteria, we detect seven and four HAEs in CO~(3--2) and dust continuum out of 22 and 19 HAEs{, respectively,} in our targeted fields (see also Fig.~\ref{fig:postage1}-\ref{fig:postage4} for a gallery of detected sources). 
We summarize flux values {for detected sources} in Table~\ref{tab:physicparam}, {and for nondetection} in Table~\ref{tab:nondet}.
The detected sources have stellar mass $>4\times10^{10}$ M$_{\odot}$, and two of them have stellar masses exceeding $\sim~10^{11}$ M$_{\odot}$ (HAE3, HAE4).

\section{Ancillary data}\label{sec:multiband}
\subsection{MOIRCS/Subaru NIR data : Mstar and SFR}\label{sec:subaru}
The stellar masses ($M_{\star}$) and SFRs of the HAEs are derived from the broad-band emissions in $J$ and $Ks$ bands and the H$\alpha$ {emissions} within the NB-filter, respectively.
The observations are executed under the seeing limited condition, i.e., $0.7^{\prime\prime}$.
Thus far, we have obtained 8 broad/intermediate/narrow-band images in the optical-to-near infrared (NIR) range {by} using Subaru, i.e., $B$, IA427, $r^{\prime}$, $z^{\prime}$, $J$, $H$, $Ks$, and NB2288 (which is called as the `CO'-filter). 
 However, we chose to use only the above three bands because the data quality (i.e., the depth and resolution) is not as good as that in longer-wavelength imaging (I. Tanaka et al., in preparation).
Further analysis to deal with such data combining data at longer wavelengths up to radio wavelengths will be presented in one of the following papers.

Since the full description of the data reduction and analysis for these observations will be presented in I. Tanaka et al. (in preparation), we present here a {only} brief summary of the derivation of physical parameters that are used throughout this paper.
The stellar mass is derived using $[J-Ks]$ color and $Ks$ magnitude and calibrated from empirical fitting between \citet{Bruzual2003} (BC03) and {the} spectral energy distribution (SED) fitting with the FAST code\footnote{http://w.astro.berkeley.edu/~mariska/FAST.html}.
The star-formation rate (SFR) is converted using {the method described in} \citet{Kennicutt2012} from the H$\alpha$ flux that is measured from the NB filter excess. 
The {\it intrinsic} star formation rate is estimated by taking into account dust extinction in the H$\alpha$ emission using {the method described in} \citet{Garn2010}, which employs mass-dependent extinction correction.
This correction method appears to hold up to z $\sim$ 1.5 (\citealt{Sobral2012, Dominguez2013, Ibar2013}) and is often used for distant galaxies (z$\sim$2) as a proxy for dust extinction (e.g., \citealt{Sobral2014}). 
We will discuss later the effect of the adopted dust correction {method} (Sec~\ref{sec:extinction}).

For massive galaxies ($M_{\star} > 10^{10}$ $M_{\odot}$), the HAEs are, in general, located near the main sequence defined at $z = 2.5$ \citep{Speagle2014, Whitaker2012} in Fig.~\ref{fig:galaxysequence}.
We also plot two SFR-$M_{\star}$ relations to follow a few studies claiming the non-linearity of the relation (e.g., \citealt{Whitaker2012, Whitaker2014, Lee2015b}). 
In this case, the slope of the star-forming sequence is flattened at the high mass-end. 
However, even if we take this effect into account ({green dashed lines in} Fig.~\ref{fig:galaxysequence}), {the} most massive HAEs are still within a scatter of the main sequence ($\sim0.3$ dex).
The outliers on the massive-end are (potential) AGNs such as HAE1 (which is the radio galaxy \objectname{4C23.56}) and HAE5 \citep{Tanaka2011}, the SFR$_{\rm H\alpha}$ {values of which} are probably overestimated owing to AGN contamination, or HAE7, which is undetected in both CO~(3--2) and 1.1 mm, expectedly have {a} low gas budget at {the} given stellar mass, and might be close to quenching or becoming passive.

Low-mass galaxies have large uncertainties in stellar masses, mainly because of large errors in photometry of both {the} Ks and J bands with {a} low S/N.
We tentatively found a signature of enhanced star formation at {a} given stellar mass that is similarly observed in other protocluster members (\citealt{Hayashi2016}; I. Tanaka et al., in preparation).
While the SFRs might be overestimated for the less massive galaxies (e.g., see Fig 8 in \citealt{Shivaei2016a}), further investigation is beyond the scope of this paper.
Since the galaxies detected in the ALMA observations are massive enough, the uncertainties in less massive galaxies would not critically affect our discussion.

Additionally, we recently obtained adoptive optics (AO)-supported $K^{\prime}$ band images with IRCS/Subaru with a resolution of $0.2^{\prime\prime}$ for several HAEs where a natural guide star is available (Y. Koyama et al., in preparation; M. Lee et al., in preparation). 
The AO images are shown in Fig.~\ref{fig:postage1}--\ref{fig:postage4} to provide some visual hints for understanding the nature of the galaxies, but a full description and detailed analysis of the observation will be presented in the following subsequent papers.
\subsection{Spitzer : MIPS 24 um}
\begin{figure*}[ht]
\includegraphics[width=0.9\textwidth, bb = 0 0 789 900]{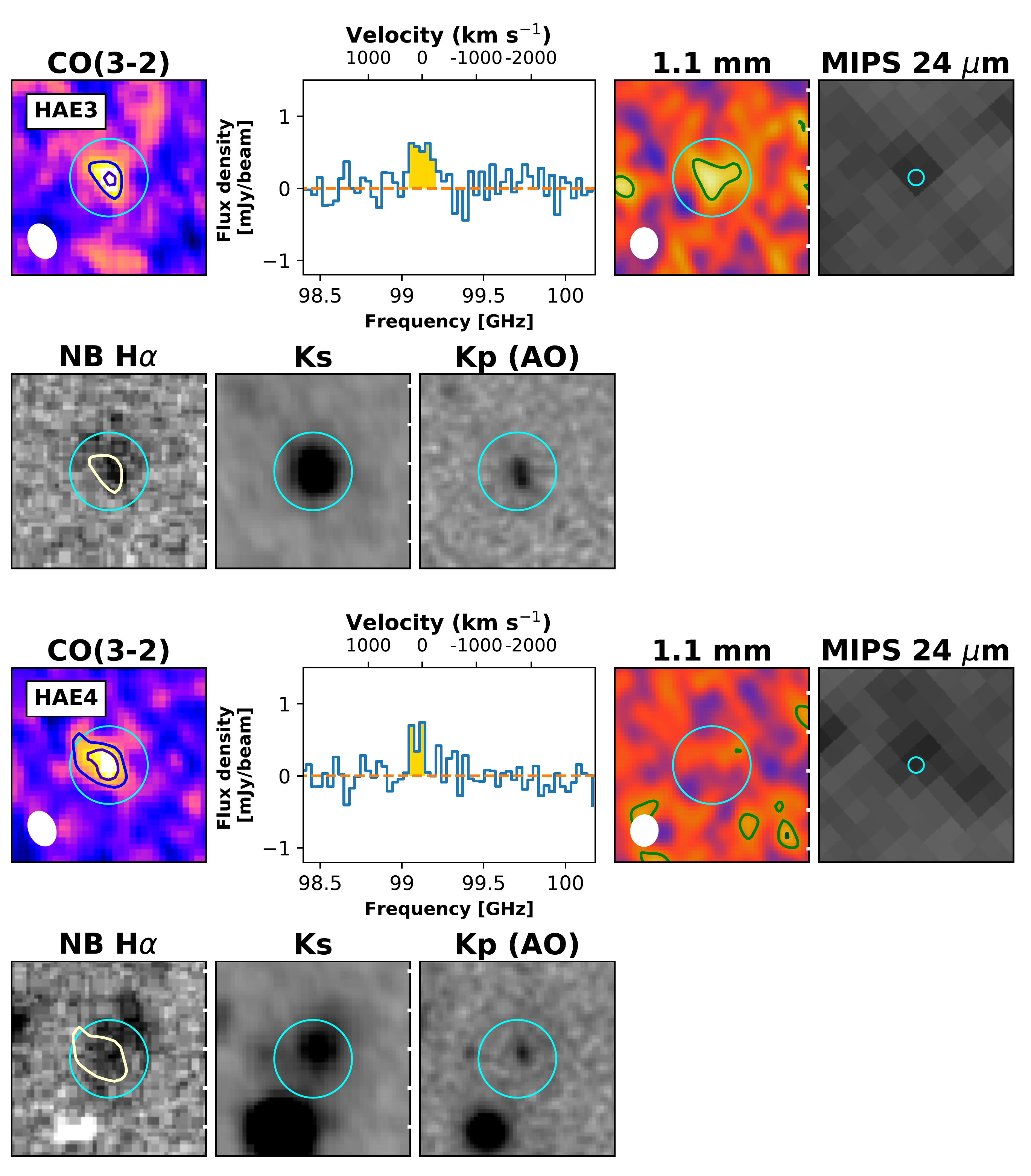}
\caption{Multi-band images of sources detected using ALMA with having either CO~(3--2) or 1.1 mm detection for HAE3 (top two rows) and HAE4 (bottom two rows). 
From left to right (upper row of each target) : CO~(3--2) integrated intensity, CO~(3--2) spectrum at the peak, 1.1 mm, MIPS 24~$\mu$m, (lower row ; continuum-subtracted NB H$\alpha$, Ks, {and} Kp (AO). 
The center of each panel is set by the CO~(3--2) peak position. 
We plot contours of CO~(3--2) and 1.1 mm emission in steps of 2$\sigma$ starting from 3$\sigma$ since the color scales of the panels are is slightly different. 
The beams of CO~(3--2) (0$^{\prime\prime}$.91$\times$0$^{\prime\prime}$.66, PA = $23.5^{\circ}$) and 1.1 mm (0$^{\prime\prime}$.78$\times$0$^{\prime\prime}$.68, PA = $0.4^{\circ}$) are shown on the bottom left.
The CO~(3--2) spectrum is shown for the range between 98.4 and 100.2 GHz into which the redshifted CO(3-2) at z$\sim$2.5 would fall.
The velocity resolution is set to 100 km s$^{-1}$ in general, but {it is set to} 30 km s$^{-1}$ for HAE5 (see Fig.~\ref{fig:postage2}). 
The yellow region of each spectrum is the integrating velocity range that delivers the highest S/N (Sec.~\ref{sec:flux}).
The 3$\sigma$ for {the} CO~(3--2) contour is also overlaid on each NB H$\alpha$ image for comparing the distribution.
In the AO images, we find compact components for the most massive galaxies among {those} detected (HAE3 and 4), while the rest are marginally visible, suggesting {the} relatively diffuse nature of the stellar component.
We also plot a cyan circle with {a radius of} 1$^{\prime\prime}${, which} is also centered on the peak position of CO(3-2), to show the scale of the panel and to point out that the counterpart at different wavelengths is located near the CO(3-2) position or within $2^{\prime\prime}$ in general (see also Appendix~\ref{app:positionerr}).
As the MIPS/Spitzer observations at 24~$\mu$m have a coarse resolution compared to those of other bands, we zoom out images to clearly show the detection.
\label{fig:postage1}
}
\end{figure*}

\begin{figure*}[ht]
\includegraphics[width=1.0\textwidth, bb = 0 0 789 900]{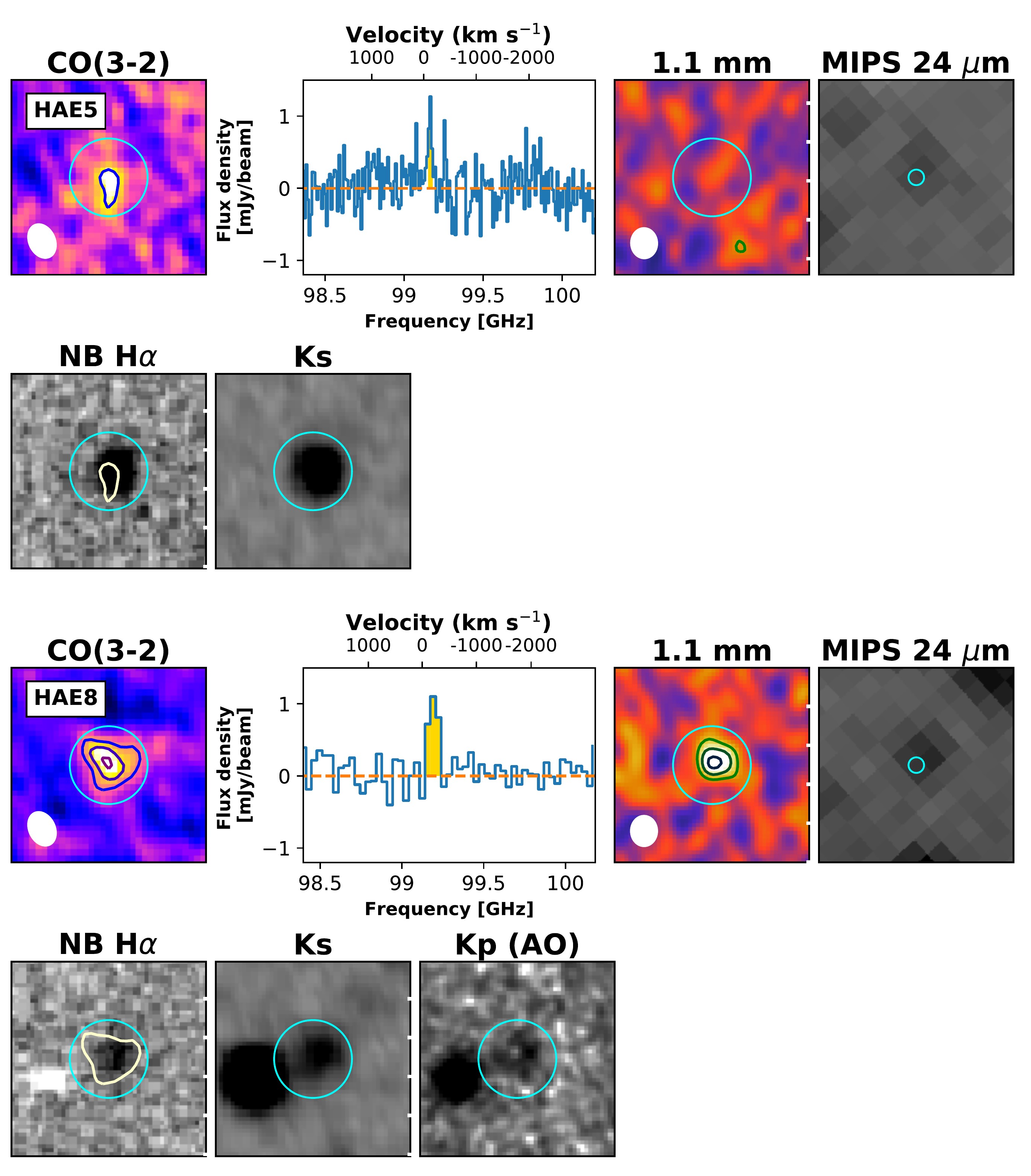}
\caption{Multi-band images for the galaxies having either CO~(3--2) or 1.1 mm detection (continued) : HAE5 (top two rows) and HAE8 (bottom two rows).
Refer {to} Fig.~\ref{fig:postage1} for the description of each panel and symbols. 
There was no coverage of the AO observation in Kp for HAE5. 
Since the line width for HAE5 is narrow (see also the text and Table~\ref{tab:physicparam}), we show the spectrum with a velocity resolution of 30 km s$^{-1}$, as opposed to other galaxies{, which are} shown with a resolution of 100 km s$^{-1}$.
\label{fig:postage2}
}
%%%%%%%%%%%%%%%%%
%%%%HAE9 and 10 gallery
%%%%%%%%%%%%%%%%%
\end{figure*}

\begin{figure*}[ht]
\includegraphics[width=1.0\textwidth, bb = 0 0 789 900]{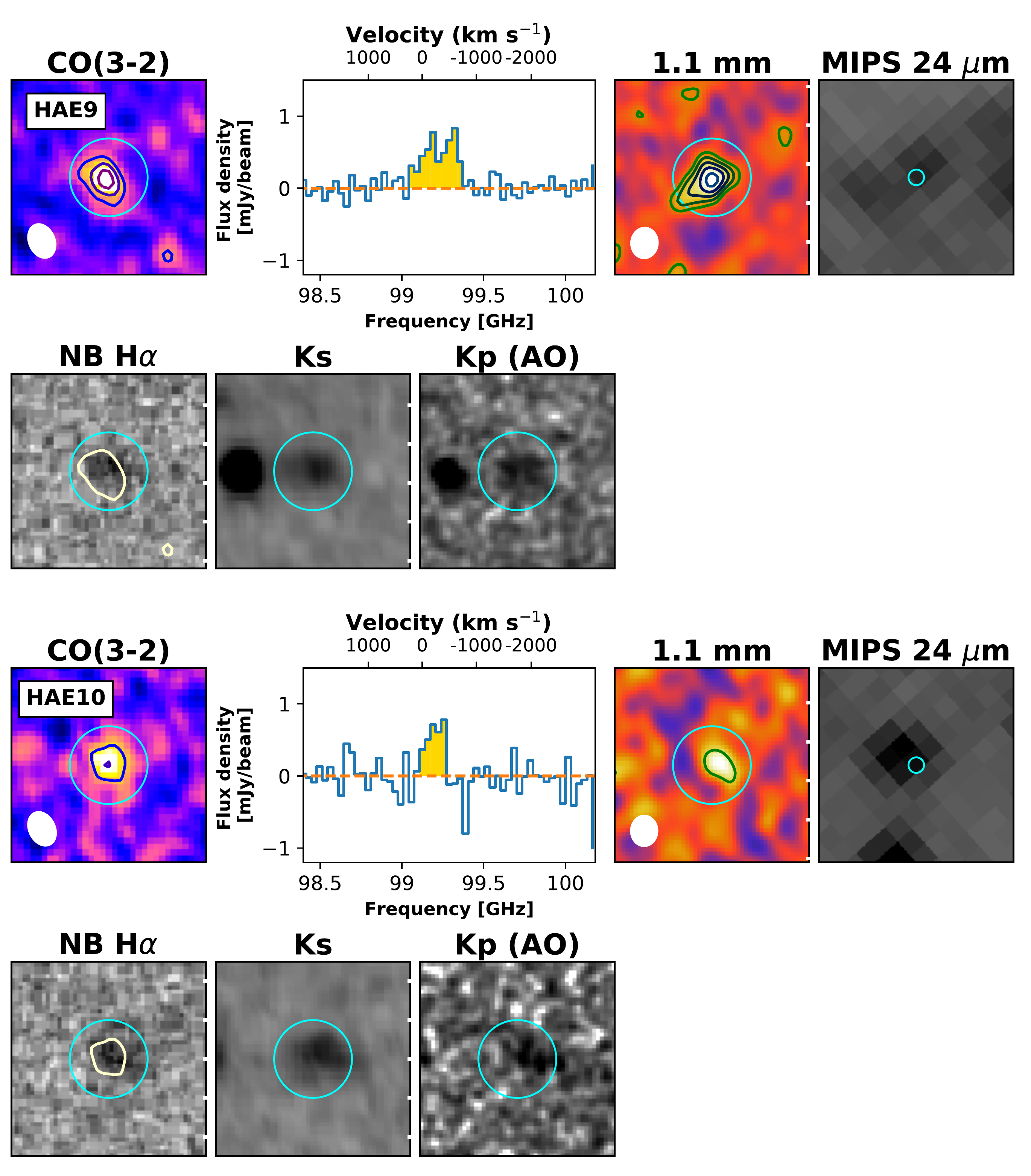}
\caption{Multi-band for the galaxies having either CO~(3--2) or 1.1 mm detection (continued) : HAE9 (top two rows) and HAE10 (bottom two rows). Refer {to} Fig.~\ref{fig:postage1} for the description of each panel and symbols.
\label{fig:postage3}
}
\end{figure*}

\begin{figure*}[ht]
\includegraphics[width=1.0\textwidth, bb = 0 0 1920 1080]{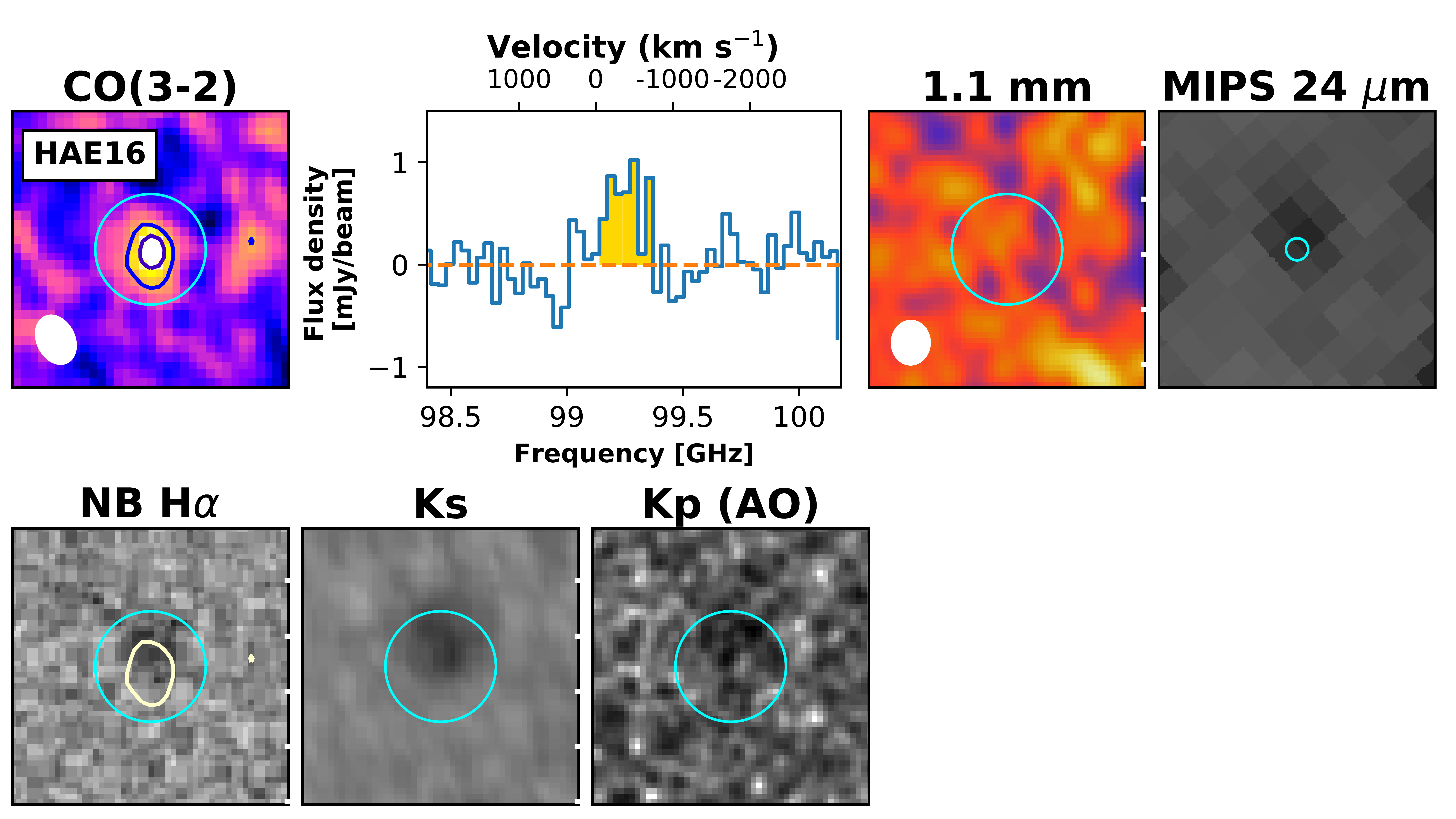}
\caption{Multi-band images for the galaxies having either CO~(3--2) or 1.1 mm detection (continued) for HAE16. Refer {to} Fig.~\ref{fig:postage1} for the description of each panel and symbols.\label{fig:postage4}
}
\end{figure*}

We also utilized archival data sets of \objectname{4C23.56} (PI: A. Stockton ; Program ID 30240) at 24~$\mu$m observed with MIPS/{\it Spitzer}, which were retrieved from the 
Spitzer Heritage Archive (SHA) interface\footnote{http://sha.ipac.caltech.edu/applications/Spitzer/SHA/}. 
We used MOPEX software package for image processing.
We present the MIPS image only to show the visual characteristics (i.e., whether a detection occurred) of the HAEs with Band~3/6 detections (Fig.~\ref{fig:postage1}--\ref{fig:postage4}).
%
%%%%%%%%%%%%%%%%%%%%%%
%%%%Figure3-6 starts: postage%%%%
%%%%-----------HAE3 and 4 gallery
%%%%%%%%%%%%%%%%%%%%%%
%%%%%%%%%%%%%%%%%
%%%%HAE5 and 8 gallery
%%%%%%%%%%%%%%%%%
%%%%
%%%%%Figure4-7end%%%%
%%%%%%%%
%
%
\section{Gas mass}\label{sec:barmass}
We measured the total gas mass from the estimated flux (Sec~\ref{sec:flux}) of dust continuum and CO~(3--2) line emission. 
In the following two sections, we address how the gas mass is estimated.
\subsection{CO~(3--2) to gas mass}\label{sec:comass}
Although the CO line emission constitutes only a fraction of the total gas content, the strategy {of} using the optically thick CO line emission in the total gas mass has been established around the time millimeter observation became available in {the} late 1980s \citep{Dickman1986, Solomon1987}.
While higher-$J$ rotational transitions of CO have large uncertainty for the unknown excitation, lower-$J$ ($J<4$) lines are a good probe for {the} total cold gas mass (e.g., \citealt{Carilli2013}), and the lines have been used for several pioneering works on high-$z$ star-forming galaxies as well as SMGs (e.g. \citealt{Magdis2012, Tacconi2013, Daddi2015})

We derived the gas mass from CO~(3--2) emission by following the prescription presented in \citet{Genzel2015}.
Provided (massive) HAEs are on the main sequence, {and} a typical conversion factor for normal star-forming galaxies or Milky-way-like galaxies, $\alpha_{\rm MW} = 4.36$ $M_{\odot} \,(\rm K\, km \, s^{-1}\, pc^2)^{-1}$, is adopted to the first order. Then the metallicity (Z) dependence of the conversion factor, i.e., $\alpha_{\rm CO\,1}$ =  $\alpha_{\rm MW} \times A(Z)${, is considered}. 
A(Z) corresponds to {the} metallicity dependence of the conversion factor calculated by taking the geometric mean of \citealt{Bolatto2013} (the equation (6) in G15) and \citealt{Genzel2012} (the equation (7) in G15).

To account for the metallicity dependence of the conversion factor, we adopted the galaxy's metallicity derived from an empirical mass-metallicity relation, as presented in \citet{Genzel2015} (equation (12a), which uses a fitting function of \citealt{Wuyts2014}). 
The adopted metallicity-dependent conversion factor is in the range of $\alpha_{\rm CO, 1}=[4.4, 5.9]$.
The reason for using the empirical relation is that we still had incomplete metallicity measurements for all the samples; only a fraction of [NII] and H$\alpha$ spectroscopic data are obtained and some have low S/N.
Within the stellar mass range of HAEs detected in CO~(3--2) and dust continuum ($4\times10^{10}<M_{\star}/M_{\odot}<2\times10^{11}$), the metallicity varies within a modest range ([8.50, 8.65]) even {if} we adopt {a} different metallicity recipe ; for example, {the one described in} \citet{Mannucci2010} would yield a value lower by $< 0.02$ dex, which results in a conversion factor that does not vary by more than a factor of 2.
It is worth noting that there might be a tendency of lower metallicity in high-$z$ overdense region (e.g., \citealt{Valentino2015}), where a pristine gas is likely being accreted from the cosmic web particularly at high redshift. 
However, this is controversial given several contradictory cases, such as higher metallicity (e.g., \citealt{Steidel2014, Shimakawa2015}), a flat mass-metallicity relation ({thus, higher metallicity in lower mass regime} e.g., \citealt{Kulas2013}), and the same {mass-metallicity relation} as fields (e.g., \citealt{Tran2015}) at z$\sim$2.
Therefore, we stick to the general comprehension of {the} stellar mass--metallicity relation.
We discuss later the validity of the choice of conversion factor in Section~\ref{sec:conversionfactor}

We use a standard luminosity (brightness temperature) line ratio between different rotational transitions of CO, i.e., CO~(1--0)-to-CO(3-2) ratio $R_{13} = 1.9$, which can be applied to {both} high-$z$ typical star-forming galaxies and SMGs (e.g., \citealt{Tacconi2008, Tacconi2013, Carilli2013, Daddi2015}). 

The gas mass is then computed as {expressed by} Eq.~\ref{eq:Mmolgas} at {a} given luminosity $L'_{{\rm CO}\, J}$ {by} using a conversion factor $\alpha_{\rm CO\, 1}$, CO $J\rightarrow J - 1$ line flux $F_{\rm CO \,J}$, source luminosity distance $D_L$ , redshift z, and observed line wavelength $\lambda_{{\rm obs} \,J}$ = $\lambda_{{\rm rest} \,J} (1 + z)$ (\citealt{Solomon1997, Bolatto2013}), where $J=3$ in our case.
 
\begin{eqnarray}\label{eq:Mmolgas}
M_{\rm gas \, ,CO} \, [M_{\odot}] &=& \alpha_{\rm CO\, 1} \times L^{\prime}_{\rm CO \,1} \nonumber \\
& = &  1.57 \times 10^9 \left( \frac{\alpha_{\rm CO \, 1} \times R_{1\, 3} }{\alpha _{MW}} \right) \nonumber \\
& & \times \left(\frac{ F_{CO \, 3}}{\rm Jy \,km \,s^{-1}} \right)  \times (1+z)^{-3} \nonumber\\
& & \times \left(\frac{ \lambda_{\rm obs\, 3}}{\rm mm} \right)^2 \times \left(\frac{D_L}{\rm Gpc}\right)^2
\end{eqnarray}

Since we aim to compare {out survey} with other high-$z$ field\footnote{We assume that the compared galaxies are in `general' fields, which may have probed {a} presumably large volume (i.e., {a} relatively wide redshift range {to cover the large scale structure}){;} thus, cosmic variance may not significantly affect {the comparison}.}  surveys based on either CO and/or dust continuum, we apply the same analysis for {the} available data set.

For {a} CO-based survey, we referred to the PHIBBS-I sample presented in \citet{Tacconi2013}\footnote{At the date of submission, the PHIBBS-2 sample was not yet available online (\citealt{Tacconi2017})}.
The PHIBBS-I galaxies are located in several fields including the Great Observatories Origins Deep Survey-North (GOODS-N) field, Q1623, Q1700, Q2343, and Extended Groth Strip (EGS) field.
This is a CO(3-2) survey of massive galaxies ($\log(M_{\star}/M_{\odot}) \geqslant 9.5$) scattered around the main-sequence star-forming galaxies between $1<z<3$.
Later, we select PHIBBS-I galaxies within the main sequence ($\pm0.3$ dex using \citealt{Whitaker2012}) at $2<z<3$ above $\log(M_{\star}/M_{\odot})>10.6$ which are unfortunately only 7 in number, and its stellar mass range is $10.6 \leqslant \log(M(M_{\star})) \leqslant 11.2$, which is almost the same stellar mass range {as} that detected in CO(3-2) (i.e., $10.6 \leqslant \log(M(M_{\star})) \leqslant 11.3$).
We apply the same gas recipe for {the} PHIBBS-I galaxies, while the values of $M_{\star}$ and SFR are simply adopted from Table~2 in \citet{Tacconi2013}, which is derived from SED fitting.

We summarize the measured CO line flux and the derived
total molecular gas masses for individual HAEs in Table~\ref{tab:physicparam}. 
The derived molecular mass ranges between $(0.3-1.9)\times 10^{11}$ M$_{\odot}$. 
The upper limit of molecular gas {mass is} set to $3~\sigma$ assuming a velocity width of $\sim300$~km~s$^{-1}$, i.e., a typical galactic disk rotation, as presented in Table~\ref{tab:nondet}. 

\subsection{1.1 mm dust to gas mass}\label{sec:dustbar}
We derive gas mass from dust continuum detection
using a method presented in \citet{Scoville2016}.
As \citet{Scoville2016} and \citet{Berta2016} have argued, the dust mass fitted with the FIR-only SED (i.e. using a SED model that is fitted only around the FIR peak with Herschel) would 
yield significant uncertainties in measuring total gas mass, 
since the flux around the peak is no longer optically thin{;}
therefore, the dust (and gas) mass {fitted} by {the} SED model is a rather luminosity-weighted value.
Therefore, we assume a dust temperature of 25 K to weigh the global gas amount as suggested in \citet{Scoville2014, Scoville2016}. 
\citet{Scoville2014, Scoville2016} derived the gas mass 
using the Rayleigh-Jeans (RJ) tail of the dust spectrum and
{by} adopting a locally calibrated luminosity-mass relation. 
The gas mass is calculated as follows :
\begin{eqnarray}\label{eq:Scovillemass}
M_{\rm gas\,, dust}\, [M_{\odot}] &=& \frac{1.78 \times 10^{10}}{(1+z)^{4.8}} \left( \frac{\Gamma_{\rm RJ}}{\Gamma_{0}}\right)^{-1} \frac{6.7 \times 10^{19}}{\alpha_{850}} \nonumber \\
& & \times \left( \frac{S_{\nu}}{\rm mJy}\right) \left( 
\frac{\nu}{\rm 353 GHz} \right)^{-3.8} \left( \frac{D_{\rm L}}{\rm Gpc}\right)^2
\end{eqnarray}
where $S_{\nu}$ is the observed dust continuum flux in mJy,
$\alpha_{850}$ is a constant for calibrating luminosity to gas mass, 
and $\frac{\Gamma_{\rm RJ}}{\Gamma_{0}}$ is the RJ correction factor with $\Gamma_0=\Gamma_{\rm RJ} (0, T_d, \nu_{850}) = 0.71$ and $\Gamma_{\rm RJ}$ given by

\begin{eqnarray}
\Gamma_{\rm RJ} (T_d, \nu_{\rm obs}, z) = \frac{h\nu_{obs} (1+z)/kT_d}{e^{h\nu_{obs} (1+z)/kT_d} -1}
\end{eqnarray}

The metallicity dependence of the dust-based calibration may not affect our discussion since the stellar mass range {of the detected sources are sufficiently large}. 
However, we note that the dust-based measurement may yield a systematically lower value than the CO-based measurements (e.g., \citealt{Genzel2015, Decarli2016b}; see also some discussions in section~\ref{sec:conversionfactor} and Fig.~\ref{fig:MmolSFR}).

The calculated results for 1.1 mm are also summarized in Table~\ref{tab:physicparam}, and images are shown in Fig.~\ref{fig:postage1}-\ref{fig:postage4}. 
We find the gas mass derived from 1.1 mm is in the range of $(0.5-1.4)\times 10^{11}$ M$_{\odot}$ for four objects.

Similarly for CO~(3--2), for the comparison with field galaxies, we referred to the study of \citet{Scoville2016}, which targeted galaxies extensively in the Cosmic Evolution Survey (COSMOS) field within the redshift range $1<z<6$. 
We also applied the same analysis for ALMA LABOCA Extended Chandra Deep Field-South Survey (ECDF-S) Submm Survey (ALESS) SMGs which are partly covered in GOODS-S. 
We particularly focus on the main-sequence SMGs within a redshift range of $2<z<3${, the stellar mass} of which is restricted to $\log(M_{\star}/M_{\odot})>10.6${;} therefore, $10.4 \leqslant \log(M(M_{\star})) \leqslant 11.7$.
The gas mass is calculated from 870~$\mu$m {as} listed in \citet{Hodge2013} (primary beam corrected flux, column 8 in Table~3) and {by} combining {it with} the information (i.e., $M_{\star}$, SFR, redshift) from another SED fitting (i.e., MAGPHYS) presented in \citet{da Cunha2015}.
The redshift of the main sequence SMGs is restricted to $z<3$ since the 870-$\mu$m flux above $z>3$ no longer traces the RJ tail, producing large uncertainties in the estimation of gas mass for the analysis of \citet{Scoville2016}. 
\subsection{Combined results of SFR vs $M_{\rm gas}$}\label{sec:combined}
The gas masses derived from different estimators of CO~(3--2) and 1.1 mm are roughly consistent with each other; three HAEs (HAE3, HAE8, and HAE9) are roughly consistent within errors, and the $M_{\rm gas, \,dust}$  of HAE10 is less than $M_{\rm gas, \,CO}$. 
The latter case might be related to the variation of dust-to-gas ratio (thus metallicity) and optically thin CO emissions, which are difficult to entangle with the given data (see Sec.~\ref{sec:conversionfactor} for discussion). 

A tension between two estimators may still exist. 
The gas mass derived from 1.1 mm is systematically smaller for all four cases, even though the sensitivity limit of {the} 1.1 mm {observations} is deeper in terms of the gas content with the prescription of \citet{Scoville2016} (see also Sec.~\ref{sec:conversionfactor}).

We will focus on the results of CO~(3--2) since the detected number is larger. 
{We} perform comparison with other surveys (those presumably in general fields), mainly the results of \citet{Genzel2015} and \citet{Tacconi2017} in which the scaling relation of gas depletion time and molecular gas fractions in general fields {was derived} from CO~(3--2) measurements.

\begin{figure}[t]
\includegraphics[width=0.5\textwidth, bb = 0 0 1024 1024]{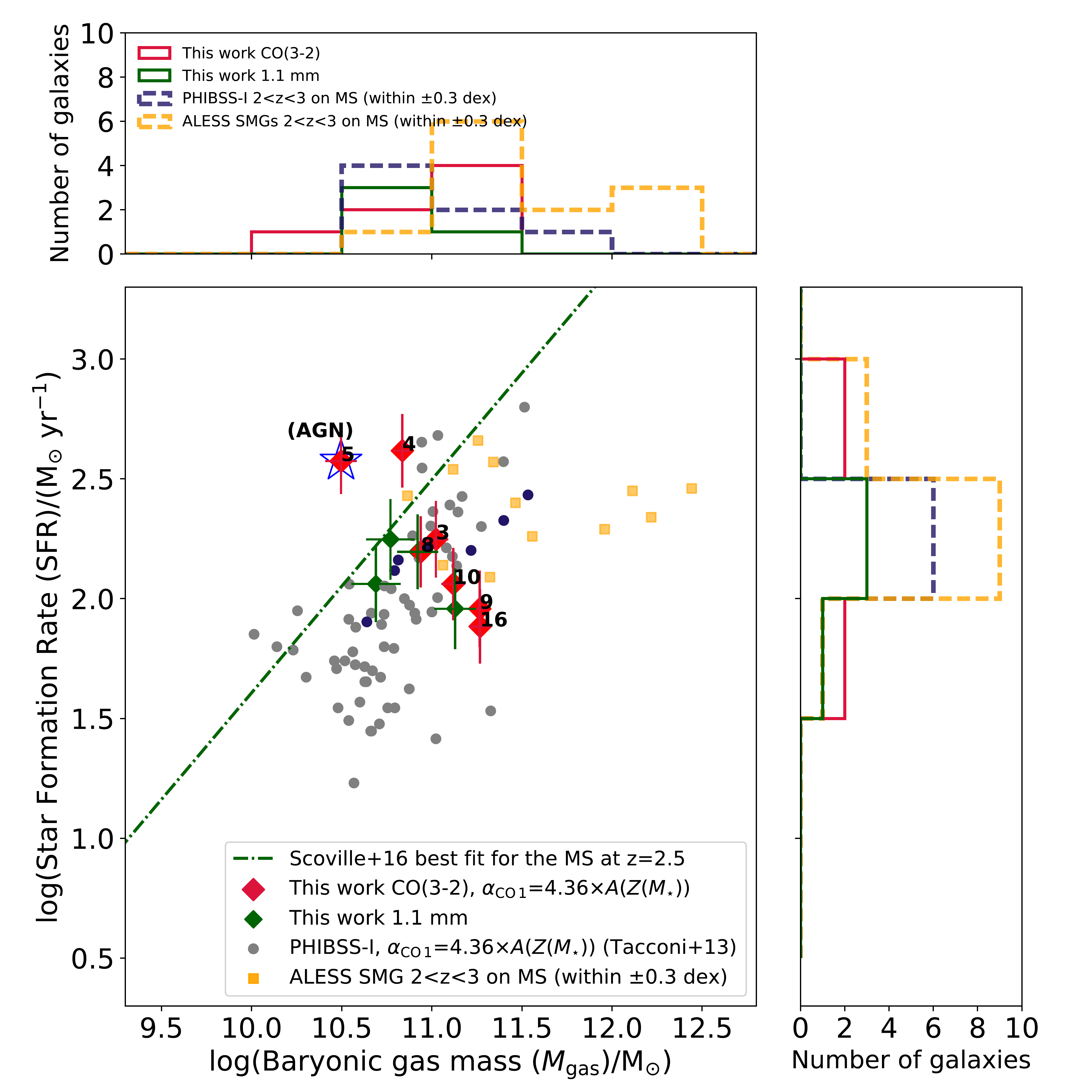}
\caption{Derived molecular mass distribution with respect to SFR. The molecular mass is derived from CO~(3--2) (red diamonds) or dust continuum (green diamonds) detection (see Section~\ref{sec:barmass} for details).
HAE5 is {indicated} with a star symbol to clarify the existence of AGN, the SFR {of which} derived from the H$\alpha$ emission may be overestimated.
We {also} plot other high-$z$ molecular and dust continuum survey results from PHIBBS-I (\citealt{Tacconi2013}), galaxies in the COSMOS field (\citealt{Scoville2016}) and ALESS (\citealt{Hodge2013, da Cunha2015}) by applying the same analysis on $M_{\rm gas}$ (but not for SFR or $M_{\star}$).
The PHIBBS-I survey (grey circles) is based on the CO(3-2) measurements for star-forming galaxies on the main sequence.
We indicated in dark blue the PHIBSS-I galaxies that are massive ($M_{\star}>4\times10^{10}$ M$_{\odot}$) on the main sequence ($\pm0.3$ dex) within $2<z<3$.
\citet{Scoville2016} (dashed green line) is based on the dust continuum (Band~7 at 870~$\mu$m) observation.
{The} ALESS survey is also observed at {the} 870~$\mu$m continuum {by} using ALMA, 
but {the observation was made} toward LESS SMGs found in the ECDF-S field. 
Yellow squares are massive ($M_{\star}>4\times10^{10}$ M$_{\odot}$) SMGs on the main sequence within $2<z<3$.
 At {a} given SFR, the gas content is roughly consistent with PHIBBS-I, while ALESS SMGs on the main sequence have {a} higher gas content, perhaps {because of} the nature of its selection.
\label{fig:MmolSFR}
}
\end{figure}

Apparently, a systematically different correlation (anti-correlation with Pearson's correlation coefficient $r=-0.85$ with {a} $p$-value {of} 0.01) between SFR and $M_{\rm gas}$ is found in protocluster members, even though the median SFE is consistent with the average value of PHIBBS samples at similar sSFR {values} ($\langle {\rm SFE}\rangle\sim$ 1.8 Gyr$^{-1}$) (Fig.~\ref{fig:MmolSFR}).
We discuss the issue further in Sec~\ref{sec:extinction} but note {here} that the apparent anti-correlation is mainly due to two populations : (i) AGN-dominated HAE5 and (ii) less massive galaxies among {the} detected {galaxies}, i.e., HAE9 and HAE16 with large velocity widths, in which the uncertainties of SFR from H$\alpha$ is expectedly larger than those in other cases.
Additionally, such anti-correlation (or no correlation) is observed in the ALESS SMGs on the main sequence with less significance ($r=-0.43$ with $p$-value=0.13).
We investigate the anti-(or no) correlation in the discussion section, and the difference might hint at an environmental effect of galaxy evolution during cluster formation.

In relation to this result, we note that all galaxies detected in CO~(3--2) have been detected in MIPS 24~$\mu$m (Fig.~\ref{fig:postage1}-\ref{fig:postage4}).
The natural correlation between the total ISM content (traced by CO~(3--2)) and star-forming activity (traced by 24~$\mu$m) (= KS relation) supports this idea.
The MIPS 24~$\mu$m emission at this redshift traces the rest-frame 7.7~$\mu$m and 6.2~$\mu$m polycyclic aromatic hydrocarbon (PAH) features (for the main sequence galaxies), and the flux can be interpreted as the SFR of the galaxy (\citealt{Lagache2004}).
However, the 24~$\mu$m flux can also be a tracer of the warm dust component heated by an AGN (\citealt{Rigby2008}).
HAE5 with a broad-line AGN signature (\citealt{Tanaka2011}) is an example that might weaken the positive correlation.
An environment that may result in a weak correlation (with large scatter) is a place of intense radiation field, for example, the (compact) galaxies with high IR luminosity (i.e., starbursts) (e.g., \citealt{Elbaz2011}).
The 24~$\mu$m flux may be also reduced in low metallicity and {a} hard radiating field (if any) (e.g., \citealt{Shivaei2016b}).
All these related factors will be further discussed in the subsequent papers.
%%%%%%%%%%%%%

\begin{figure}[th]
\includegraphics[width=0.48\textwidth, bb = 0 0 1024 1000]{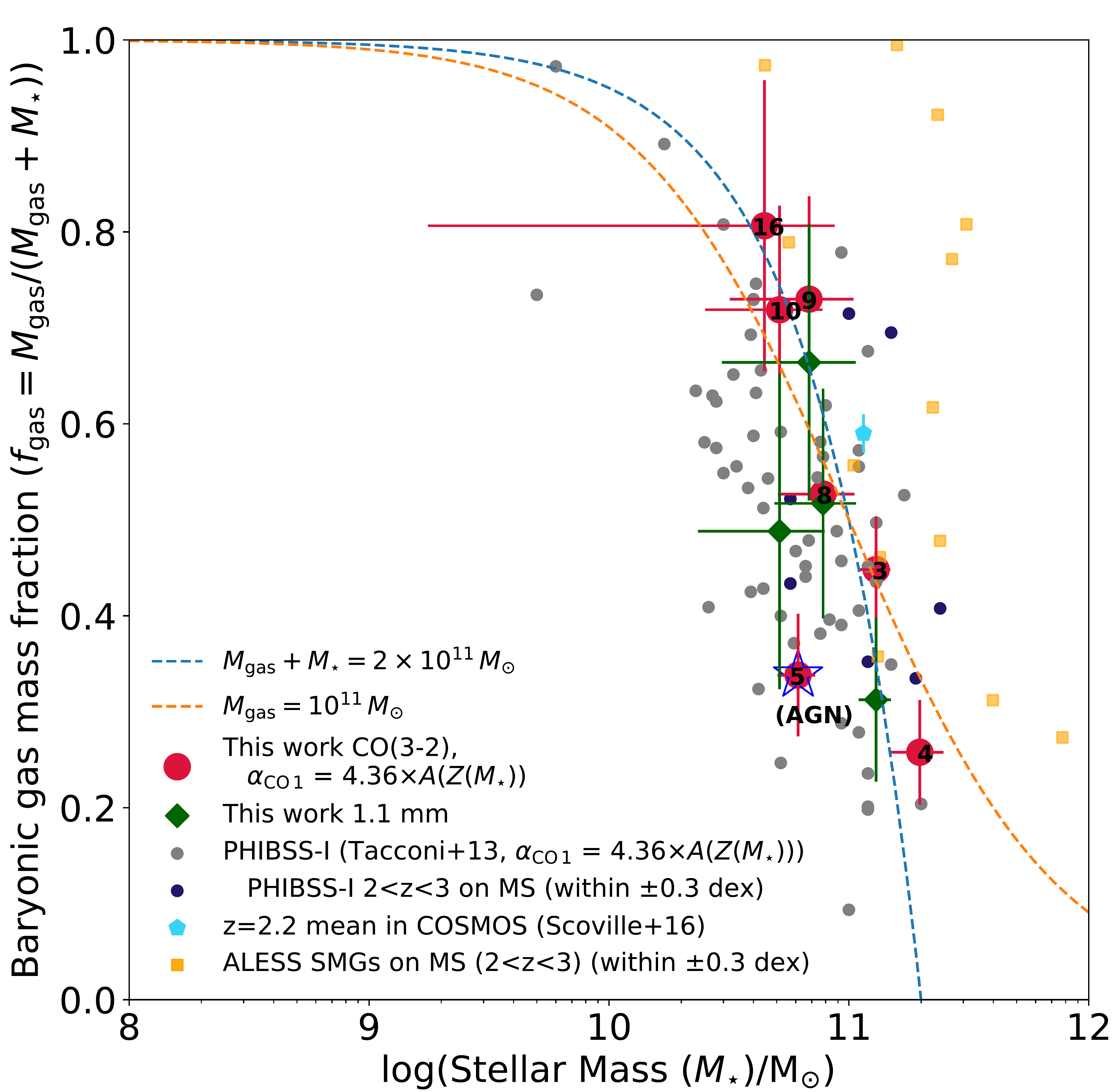}
\caption{Gas fraction ($f_{\rm gas}=M_{\rm gas}/(M_{\rm gas}+M_{\star}$)) as a function of stellar mass ($M_{\star}$). The same color scheme {is} used in Fig.~\ref{fig:MmolSFR} for HAEs in the protocluster, PHIBBS-I, and ALESS SMGs on the main sequence. 
The result from \citet{Scoville2016} for $\langle z\rangle$=2.2 mean is plotted with the cyan pentagon. At given stellar mass the protocluster members have similar gas fraction distributions with those in general fields.\label{fig:fgasMstar}
}
\end{figure}

\begin{figure}[th]
\includegraphics[width=0.5\textwidth, bb = 0 0 1024 950]{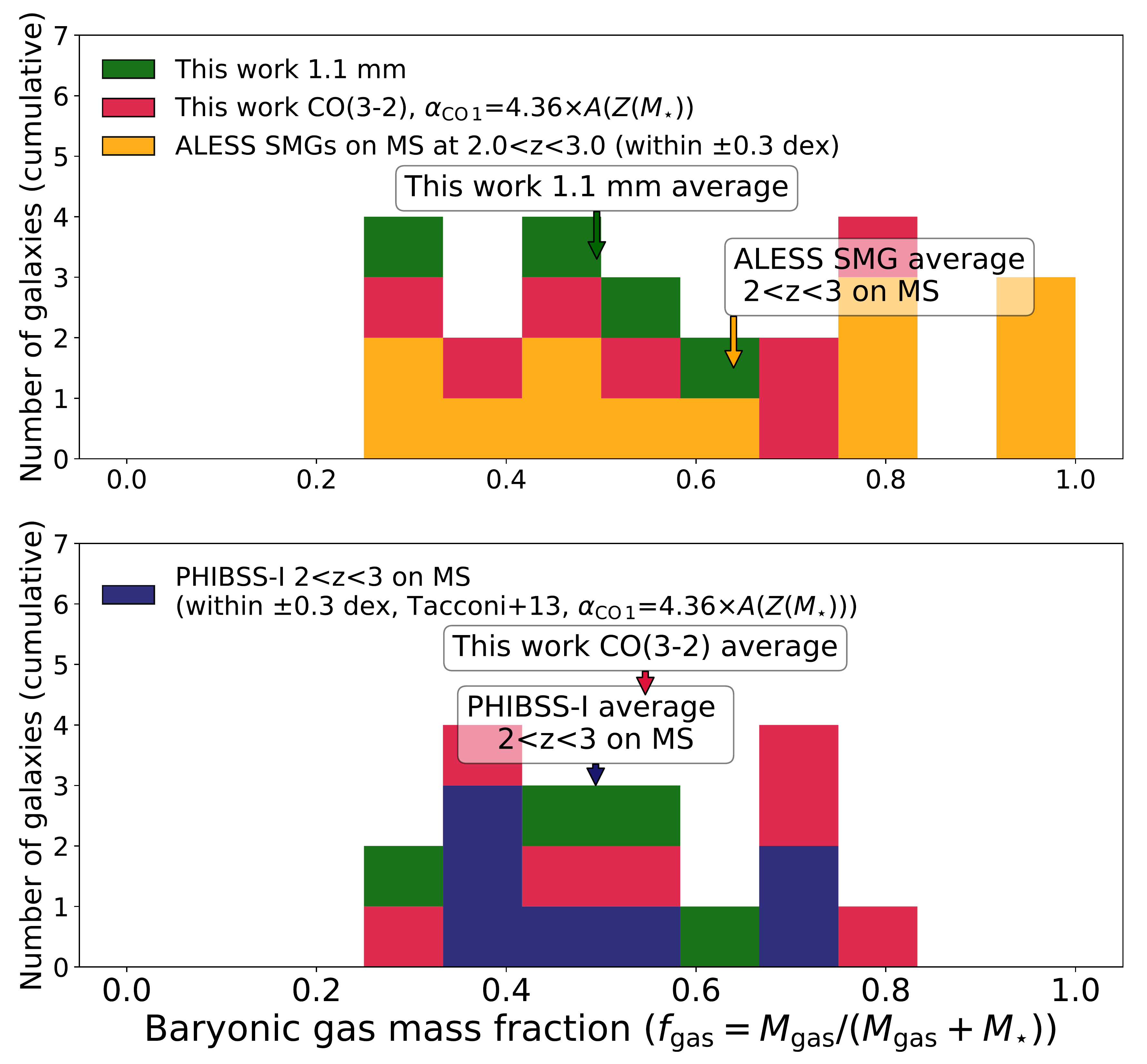}
\caption{Histogram of gas mass fraction for the massive ($>4\times10^{10}\, M_{\odot}$) main-sequence galaxies at $2<z<3$. 
We plot the distribution of the ALESS SMGs on the top panel and {that of} PHIBSS-I on the bottom. 
In general, the distribution of gas fraction (scatter $\sigma_f=0.20$ for CO~(3--2) and $\sigma_f=0.12$ for 1.1 mm) and the average ($\langle f_{\rm gas}\rangle = 0.55\pm 0.07$ for CO~(3--2) and $0.50\pm0.06$ for 1.1 mm) of the protocluster galaxies are consistent with PHIBBS-I ($\langle f_{\rm gas}\rangle = 0.49 \pm 0.05$, $\sigma_f=0.14$), but with a slightly larger scatter ($\sigma_f=0.24$) and average ($\langle f_{\rm gas}\rangle = 0.64\pm0.07$) for ALESS SMGs on the main sequence. 
SMGs have {a} slightly higher value, perhaps because of the selection effect.\label{fig:fgashist}
}
\end{figure}
%%%%%%%%%%%%

\subsection{Gas fraction}\label{sec:gasfraction}
We calculated the gas fraction ($f_{\rm gas}$ = M$_{\rm gas}$/(M$_{\rm gas}$+M$_{\star}$)) from the estimated gas mass and the stellar mass. 
The average value of the gas fraction is $\langle f_{\rm gas}\rangle$ = $0.55\pm0.07$ for CO~(3--2) and $0.50\pm0.06$ for 1.1 mm, and the values are roughly consistent with each other.

We found that the gas fraction strongly depends on the stellar mass, as in {the} PHIBBS-I sample (Fig.~\ref{fig:fgasMstar}). 
Such {a} mass dependency of gas fraction may be regulated by the mass dependent feedback and/or the gas accretion efficiency as previously addressed in cosmological hydrodynamical simulations (e.g., Fig.11 in \citealt{Tacconi2013} which uses \citealt{Dave2011}). 
We will additionally discuss in the following section the potential role of the environment in this picture.
However, we note that Fig.~\ref{fig:fgasMstar} is also consistent with the gas mass fraction spanning the entire range, just from the stochastic nature of inflow and star formation ; after all, the gas depletion time is short for any coherent evolutionary scenario.

We compared the above results with PHIBSS-I and ALESS SMGs particularly for those on the massive ($>4\times10^{10}\, M_{\odot}$) main sequence at $2<z<3$. 
By restricting galaxies in PHIBBS-I, which results in only 7 galaxies for comparison, we find {that} the average gas fraction does not differ ($\langle f_{\rm gas, PHIBBS-I, MS}\rangle$ = 0.49$\pm$0.05) ({bottom of} Fig.~\ref{fig:fgashist}).
Other studies on the main-sequence galaxies have revealed similar results (e.g., \citealt{Magdis2012, Saintonge2013, Sargent2014, Scoville2014, Scoville2016, Decarli2016a, Schinnerer2016}).
ALESS SMGs on the massive main sequence at $2<z<3$ (total number of 13) appear to have a slightly higher mean value ($\langle f_{\rm gas}\rangle$ = 0.64$\pm$0.07) (Fig.~\ref{fig:fgashist} top) but is nevertheless consistent within an error.

\section{Discussions}\label{sec:discussions}
\subsection{Reasons for unexpected non-detection}\label{sec:nondetection}
Out of {the} seven CO(3-2) detections, only four have 1.1 mm counterparts. 
We could not detect 1.1 mm emission for HAE4, HAE5 and HAE16. 
Massive galaxies with $M_{\star}>10^{11}\, M_{\odot}$ (HAE1, HAE2) that have high SFRs from the H$\alpha$ emission and {other} galaxies with high SFRs (HAE12, HAE13) are not detected in either the CO~(3--2) or 1.1 mm emissions, as opposed to our expectation that these galaxies would be detected if the normal KS-relation apply. Further, recent observations reported the detection of the massive main-sequence galaxies (e.g., \citealt{Decarli2016b, Tadaki2016}).
There are several reasons that may apply for the non-detection.
\begin{itemize}
\item AGN-dominated galaxies (HAE1 (radio galaxy 4C23.56) and HAE5) :
although the H$\alpha$ emission detected by the NB filter may have a significant contribution from AGN so that intrinsic SFR may be smaller than the estimated {SFR},
the AGN-dominated galaxies may be intrinsically gas-poor systems because of the AGN feedback, i.e., energetic outflows blowing out the gas content (e.g., \citealt{Cicone2014}).
HAE5, detected only with CO(3-2), has one of the lowest gas contents and gas fractions.
Since the radio galaxy has gigantic bipolar radio lobes associated with the X-ray emissions (\citealt{Blundell2011}), another possibility is the lack of {a} `cold' phase gas, such as that traced by cold dust (i.e., $T_{\rm d}$=25 K) and low-J CO emissions that we observed, owing to a strong radiation field heated by the central AGN.
\item Intrinsically smaller $M_{\star}$ and SFR (HAE2) :
from our newly obtained AO-data, we found {that} the galaxy may be gravitationally lensed. 
The intrinsic stellar mass (and SFR) may be much less than expected from the seeing-limited data (I. Tanaka et al. in preparation).
\item Extended low surface brightness dust component (HAE4)? : 
the non-detection in 1.1 mm with CO(3-2) detection might suggest a lower surface brightness in the dust continuum, which is also discussed in \citet{Decarli2016b} for a galaxy that has no dust but {is detected in} CO. 
Because HAE4 has greatly extended H$\alpha$ emission compared to CO (see Fig.~\ref{fig:postage1}), the dust might also {be} extended and diffuse. 
It is unlikely from a general point of view, however, that local U/LIRGs as high-z analogs (in terms of IR-luminosity) have a compact dust component with high surface brightness (e.g., \citealt{Sakamoto2013, Saito2015}) compared to CO emissions. 
Future observation is necessary to confirm such populations.
\item The lack of sensitivity (HAE16) : 
HAE16 is observed at the edge of the FoV at Band~6, and the sensitivity was not sufficient to detect dust continuum, given the CO~(3--2) detection.
\item Lower metallicity for low stellar mass galaxies? : 
HAE12\footnote{Additionally, we note the galaxy is in the close vicinity of the radio galaxy (offset $\sim$ 25 kpc physical {size}). 
The galaxy might have encountered a strong feedback from the AGN, {for} which we need additional observations.} and HAE13 have high SFR but in the relatively lower mass ($<10^{10}\, M_{\odot}$) regime. 
The gas may be CO-dark in {terms of} the effect of photodissociation in the low-metallicity regime. 
The dust-based calibration might be no longer valid. 
Otherwise, they are gas-poor systems with high SFE.
\end{itemize}

All the potential cases we have listed have to be checked with future {observations with }increased depth and higher resolution to confirm the diversity of cold gas properties of the protocluster members.

Our findings also suggest caution regarding general expectations for the main-sequence galaxies. 
With the variety of potential reasons for unexpected non-detection of the protocluster galaxies on the main sequence, `some universality' of the main sequence may have to be carefully re-checked through observations. 
There is a wide range of gas content and SFR with different masses.

\subsection{Additional adjustment in dust extinction?}\label{sec:extinction}
Since we adopt only mass-dependent extinction correction using \citet{Garn2010}, we need to carefully consider whether the corrected SFR from H$\alpha$ is accurate.

Considering the averaged value of a galaxy population as a whole, we find {that} the correction method appears to be adoptable for ALMA detected galaxies. 
We have tested with other results derived using an extinction-free radio flux (e.g., \citealt{Kennicutt2012}) with Jansky Very Large Array (JVLA) observations at 3 GHz (10 cm) (\citealt{Lee2015a}; M. Lee et al., in preparation).
We find {that} the difference is within a few factors ($\lesssim4\times$) between the SFRs adopted in this paper and that derived from the radio flux, suggesting that they are {\it not extremely (i.e., $A_{\rm v}\gg5$}) obscured cases that relocate a galaxy well above the main sequence (i.e., $\geqslant0.6$ dex) but {are} moderately dusty.
Exceptional cases are radio-loud AGNs in which the radio flux overestimates SFR owing to the increased contribution of non-thermal synchrotron emission from the AGN, which no longer traces star formation activities of a galaxy.
The differences of individual galaxies cancelled out, and star-formation rate is, on average, roughly consistent with each other.

Nevertheless, we need to pay careful attention, given the limited number of detections with limited range ($\sim$an order of magnitude) of the parameter space (i.e., $M_{\star}$, SFR, $M_{\rm gas}$).
Particularly, the radio measurements show the SFRs of HAE16 and HAE9 (those with the highest $f_{\rm gas}$) to be $\sim3-4$ times larger than SFR$_{\rm H_{\alpha}}$ (corrected).
They will be re-located slightly above or on the upper edge of the main sequence (HAE16 : log(sSFR$_{\rm JVLA}$/sSFR(ms)) $\sim$ 0.4 dex; HAE9 : log(sSFR$_{\rm JVLA}$/sSFR(ms)) $\sim$ 0.3 dex) since they were on the lower edge of the main sequence with H$\alpha$-based measurements (see Fig.~\ref{fig:galaxysequence}).
If we adopt the radio measurements instead, with higher SFR for HAE9 and 16, the apparent `anti-correlation' between SFR and $M_{\rm gas}$ observed in Fig.~\ref{fig:MmolSFR} becomes less significant, although it still exists. 

The radio observations are also limited by the detection number and have a significant scatter with the uncertainty in the radio spectral index.
Further investigation will be conducted in future studies and is beyond the scope of this paper. 
Therefore, our best conjecture for the intrinsic star-formation rate within this paper is the use of SFR$_{H\alpha}$ (corrected) while considering the potential uncertainty for the extinction correction.

\subsection{Validity of using Galactic conversion factor}\label{sec:conversionfactor}
Many studies of typical star forming galaxies, particularly in the high-mass region where the metallicity dependence is low, have adopted the ``Galactic" CO(1-0)-to-H$_2$ conversion factor (e.g., \citealt{Dickman1986, Daddi2008, Daddi2010, Tacconi2013, Genzel2015}), 
and the U/LIRG-like conversion factor $\alpha_{\rm CO} = 0.8$ $M_{\odot} \,(\rm K\, km \, s^{-1}\, pc^2)^{-1}$ is used for galaxies above the main sequence at {a} high redshift (e.g., \citealt{Solomon2005, Yun2015}). 

Our findings suggest that the use of $\alpha_{\rm CO} = 4.36 $ (M$_{\odot}$ (K km s$^{-1}$ pc$^2$)$^{-1}$) as the first order is favorable for the protocluster galaxies on the main sequence. 
It renders the gas masses derived from different calibrations, i.e., CO(3-2) and dust continuum, consistent with each other within errors.
The U/LIRG-like conversion factor yields larger inconsistencies between different estimators since the gas mass derived from CO is smaller than dust measurement {by a factor of 2-5}.
If it were applicable, this would require a higher dust temperature higher by a factor of 2-5, since the gas mass recipe of \citet{Scoville2014, Scoville2016}, the gas mass {is} inversely proportional to the dust temperature (i.e., higher RJ correction factor $\Gamma_{\rm RJ}$ with increasing dust temperature). 
Such {a} high dust temperature is unlikely at {the} observed resolution. The observed resolution and the size {(measured when it is resolved)} can probe the average temperature of the galaxy as a whole. 
Otherwise, it would be extremely compact ($<1$ kpc) in size.

Provided the small number of detections, we may be able to examine further the validity of adopting {the} ``Galactic'' conversion factor, when (i) larger samples (with a larger mass range) and (ii) different measurements (e.g., multiple-$J$ CO line or {a} simpler optically thin line such as [CI]) are available.

Before closing this section, we list several considerations for the adoption of the  conversion factor.
\begin{itemize}
\item Large line width : we found {that} more than two-thirds of galaxies have velocity widths 
$> 300$ km s$^{-1}$, i.e., very disturbed similar to on-going mergers observed in local U/LIRG (M. Lee et al., in preparation). 
In this case, CO emission might be optically thin, requiring the conversion factor to be lower than the assumed value.
We note, however, that in \citet{Daddi2010}, six BzK galaxies detected with CO(2-1) have large FWHM ($>500$ km s$^{-1}$), and the authors used the ``Galactic" value ; one of the six galaxies is possibly a rotating disk in {the} velocity-position diagram, while the others cannot be directly tested to determine whether they are rotating.
\item Uncertainties in the contribution of atomic content : we assumed that the molecular gas is dominant in high-$z$ galaxies since the mean H$_2$ column densities and ISM pressure are expectedly higher than the local values (e.g., \citealt{Obreschkow2009b}).
Furthermore, as a protocluster is similar to a group-like environment (e.g. \citealt{Toshikawa2014}), shock heating might prevent HI gas from accreting onto a galaxy (e.g., \citealt{Appleton2013}) or the neutral gas may be stripped while galaxies form a common halo (e.g., \citealt{Verdes-Montenegro2001}) leading to a lower HI content compared to that of the fields.
In addition, the gas accreted particularly onto massive galaxies around high-z overdensities may be recycled gas (\citealt{Emonts2016}).

\item Gas mass from CO {always higher} than that derived from 1.1 mm : 
we find a systemic offset between CO-based and dust-based measurements, and a similar trend was previously reported from several studies (e.g., \citealt{Genzel2015, Decarli2016b}).
\citet{Genzel2015} argued that referring {to the} true dust temperature (at least from two bands) and correcting for metallicity would improve the inconsistency.
It might also be due to the more extended and diffuse nature in 1.1 mm, where the extended emission below the surface brightness limit is missed (see also Sec.~\ref{sec:nondetection}). 

\end{itemize}
Additionally, we have discussed the validity of {the} ``Galactic" conversion factor from the dynamical mass point of view that includes large uncertainties without measuring the size and inclination (see Appendix~\ref{app:dynamical}). We will revisit this issue with future observations.
\subsection{Gas content in a protocluster}\label{sec:bargas}
We find {that} our protocluster members have on average, similar gas fractions of main-sequence {field} galaxies (see Fig.~\ref{fig:fgasMstar} and ~\ref{fig:fgashist}). 
The ALESS SMGs on the main sequence may have slightly higher gas fractions, but {they are} consistent within errors.
Since ALESS SMGs were pre-selected by their dusty nature, i.e., bright SMGs in the LESS sample (\citealt{Hodge2013}), gas-rich main-sequence galaxies may have been selectively chosen.
In either case, the gas fractions for all of the high-$z$ galaxies are higher than the local value (of star forming galaxies, $f_{\rm gas}\sim 0.08$) at {a} given stellar mass (e.g., \citealt{Saintonge2013, Tacconi2013}).
%
%%%%%%%%%%%%
\begin{figure}[tb]
\includegraphics[width=0.5\textwidth, bb = 0 0 1024 950]{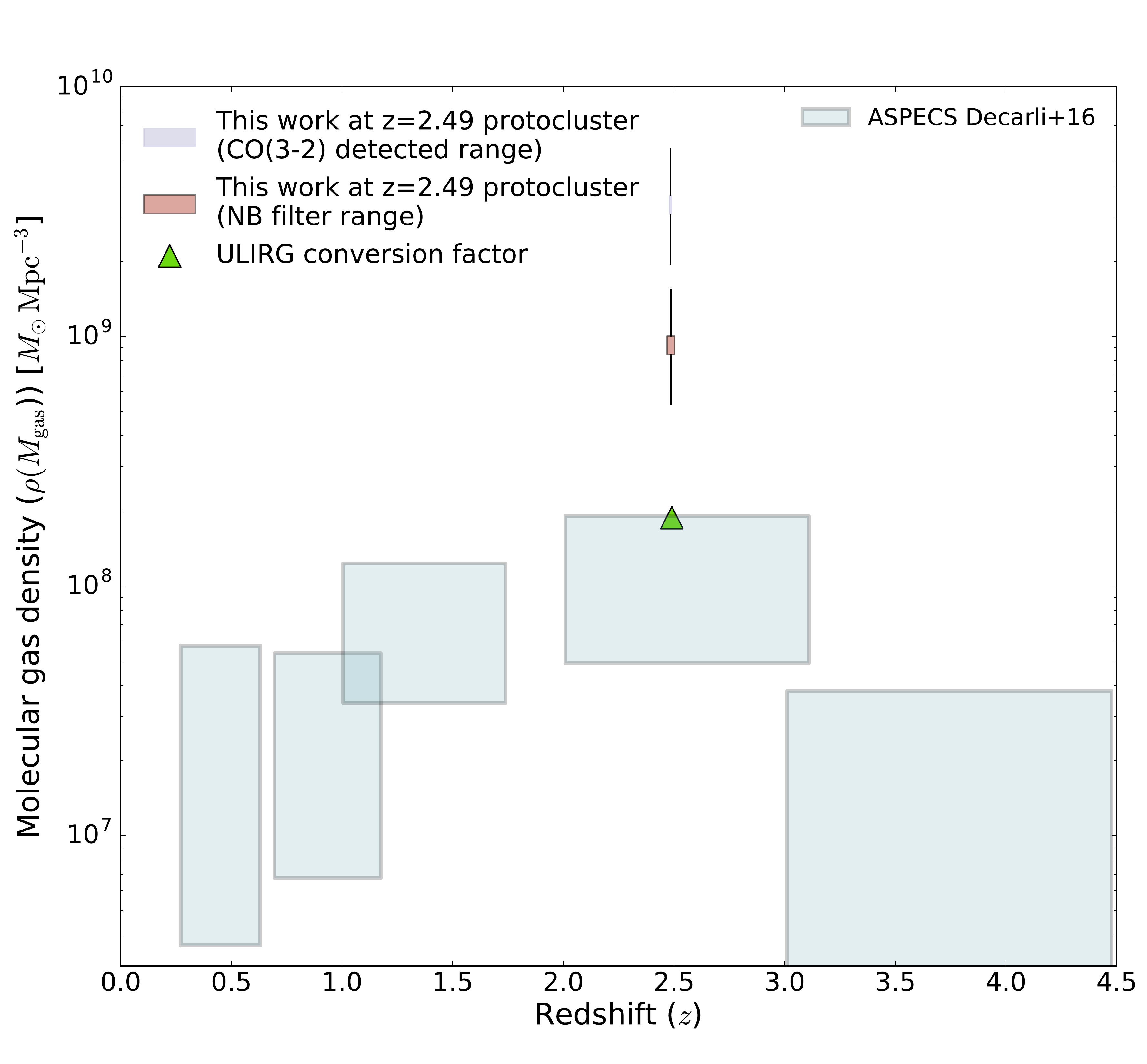}
\caption{Cosmic gas density for the z=2.49 protocluster (this work) overlaid on the recent ALMA studies of general field, HUDF-S (\citealt{Decarli2016a}) (see the text on details of the calculation to match both results). We plot three different estimations (i) using CO(3-2) redshift range ($\Delta z\sim0.01$), (ii) using NB filter redshift range ($\Delta z\sim0.03$) and (iii) applying U/LIRG conversion factor ($\alpha_{\rm CO}=0.8$ for the case (ii). The black error bar is estimated by taking into account Poisson uncertainties (\citealt{Gehrels1986}).\label{fig:gasdensity}
}
\end{figure}
%%%%%--figure ends
%%%%%%%%%%%%

We estimate the cosmic gas density of the protocluster (Fig.~\ref{fig:gasdensity}). 
The survey area is 14 comoving Mpc$^2$, and we adopted a $\sim$20\% sensitivity region (a radius of 37$^{\prime\prime}$) with our 4-pointing observations. 
If $\Delta$z is restricted only to the {sources detected in} CO~(3--2), which results in the range of $2.478<z<2.487$ ($\sim11$ comoving Mpc), then the cosmic gas density is estimated {as} $\rho_{\rm gas, 4C23.56}\sim 5 \times 10^{9} \,M_{\odot}\, {\rm Mpc}^{-3}$. 

This is $\sim22\times$ higher than the upper limit of the general field, i.e., HUDF at $z=\langle 2.6\rangle$ (\citealt{Decarli2016a}) or other previous surveys (\citealt{Walter2014, Keating2016}) and any other models 
\\(e.g. \citealt{Obreschkow2009a, Lagos2011, Sargent2014}). 
Note that we applied the same $R_{13}=2.38$ and the uniform conversion factor $\alpha_{CO\,1}=3.6$ to compare with the result of \citet{Decarli2016a}. 
This {effectively} changes the total value of $\rho_{\rm gas, 4C23.56}${ by $\sim$15\%}.

Provided a recent simulation with an expected size of the protocluster (\citealt{Chiang2013}), we could perform a more conservative derivation assuming a wider redshift range. 
We performed calculations by assuming the line-of-sight distance of the protocluster to be set by the narrow-band filter coverage ($\Delta z\sim0.03$, $\sim$40 comoving Mpc). 
The gas density becomes $1 \times 10^{9} \,M_{\odot}\, {\rm Mpc}^{-3}$, which is still a factor of six higher than the result of general fields.
A more conservative method is to derive the gas density by applying {a} U/LIRG-like conversion factor for all detected sources, lowering the gas density by a factor of 4.5, which can be regarded as the {\it lower limit} {of the gas density of the protocluster,} close to the upper limit of the general field.

Although it may not be fair to compare our results with the results of Subaru/MOIRCS, which has a different survey size ($\sim3000$ comoving Mpc$^{3}$), 
we note that protocluster 4C23.56 is also an order of magnitude higher in the cosmic star-formation-rate density (SFRD) and (3-9) times higher in the stellar mass density ($\rho_{\star}$) compared to the results presented in \citet{Madau2014} with the detection of 25 HAEs.

We barely infer the causality of these observational results. 
The higher gas density may simply be due to the higher number density of the galaxies at a given volume, which can also be inferred from the high SFRD and $\rho_{\star}$. 
Alternatively, the reason for the galaxy overdensity in the protocluster might be the higher gas density within the volume.
The former case can be simply explained by the number density of HAEs of protocluster being threefold higher than that of the field (\citealt{Tanaka2011}) and the fact that the average gas fraction is similar to the field.

However, we note that the estimated gas density is the lower limit since we only perform calculations for the detected sources.
It is uncertain whether galaxies at a lower mass regime have a larger amount of gas, and this issue cannot be clarified with the current method (or current calibration).
Nevertheless, we did find an extremely large amount of gas ($f_{\rm gas}>0.7$) in three of the galaxies, HAE9, HAE10 and HAE16, which are in the lower massive bin in our detected sample.
While noting that statistical significance is not sufficiently high to reject the opposing case given the size, the scatter of the gas fraction (observed in CO~(3--2)) is 13\% higher than that of the PHIBBS-I sample.
Further higher sensitivity observations or another tracer of the so-called CO-dark gas tracers would clarify this issue.

%%%%%--figure starts
%%%%Figure start%%%%%
\begin{figure*}[t]
\centering
\includegraphics[width=0.8\textwidth, bb = 0 0 1024 1024]{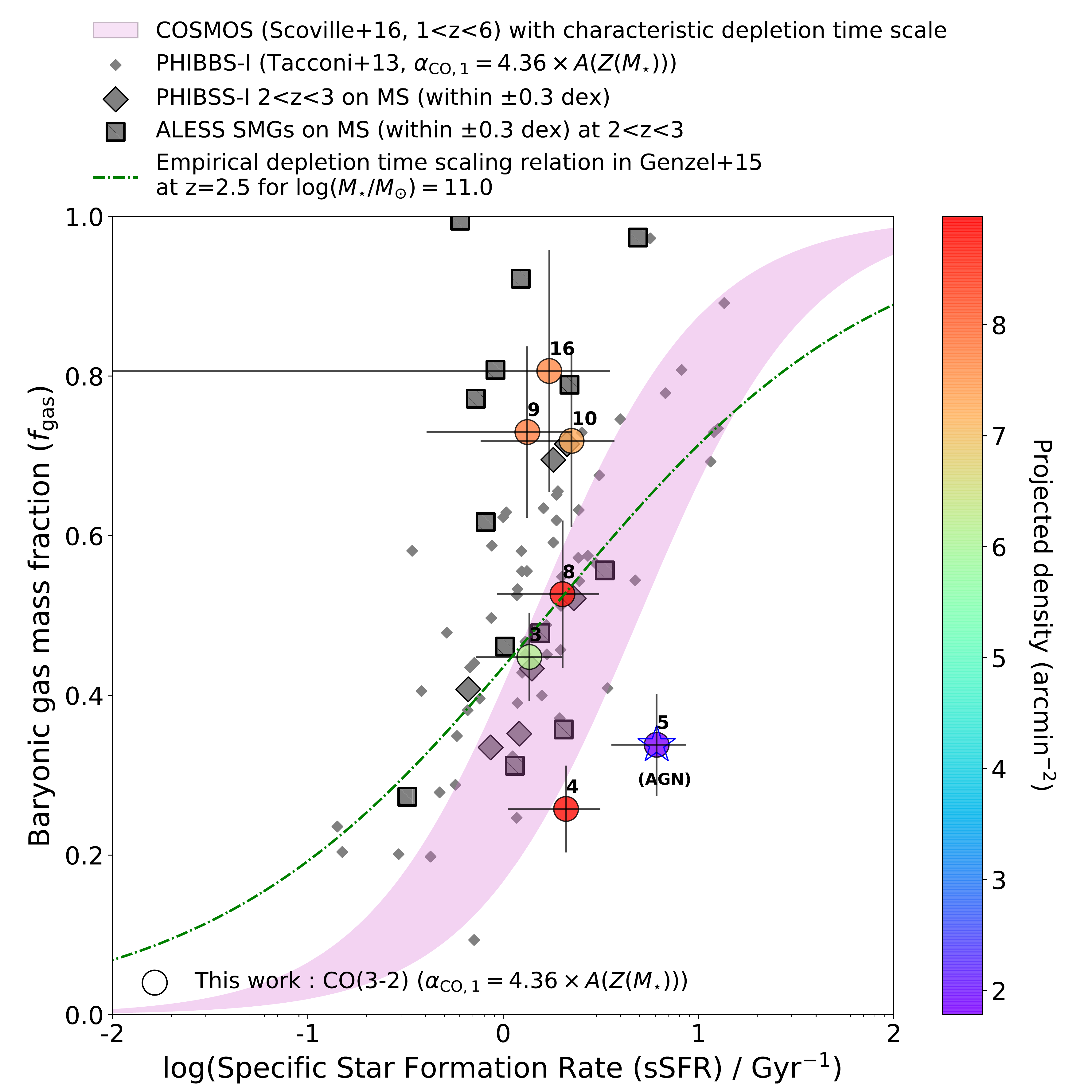}
\caption{Gas fraction as a function of specific star-formation rate. 
The color bar and numbers next to it show the relative surface number density of galaxies in arcmin$^{-1}$. 
The surface number density is estimated by the method described in the caption of Fig.~\ref{fig:fovs} and in the text. 
We plot an empirical model of {the} depletion time scaling relation described in \citet{Genzel2015} by fixing the redshift at $z=2.5$ and the stellar mass $M_{\star}=1\times10^{11}$ $M_{\odot}$. For comparison, the massive ($>4\times10^{10}\, M_{\odot}$) main-sequence galaxies at $2<z<3$ are plotted from PHIBBS-I (grey diamonds) and ALESS (grey hatched squares). 
We plot the results of \citet{Scoville2016} {by} using the characteristic depletion time of $(2-7)\times10^8$ yrs. 
A star symbol indicates the existence of AGN.\label{fig:fgas-sSFR}
}
\end{figure*}
%
%%%%%%
%%%%%Figure start%%%%%
\begin{figure*}[th]
\centering
\includegraphics[width=0.8\textwidth, bb = 0 0 1024 950]{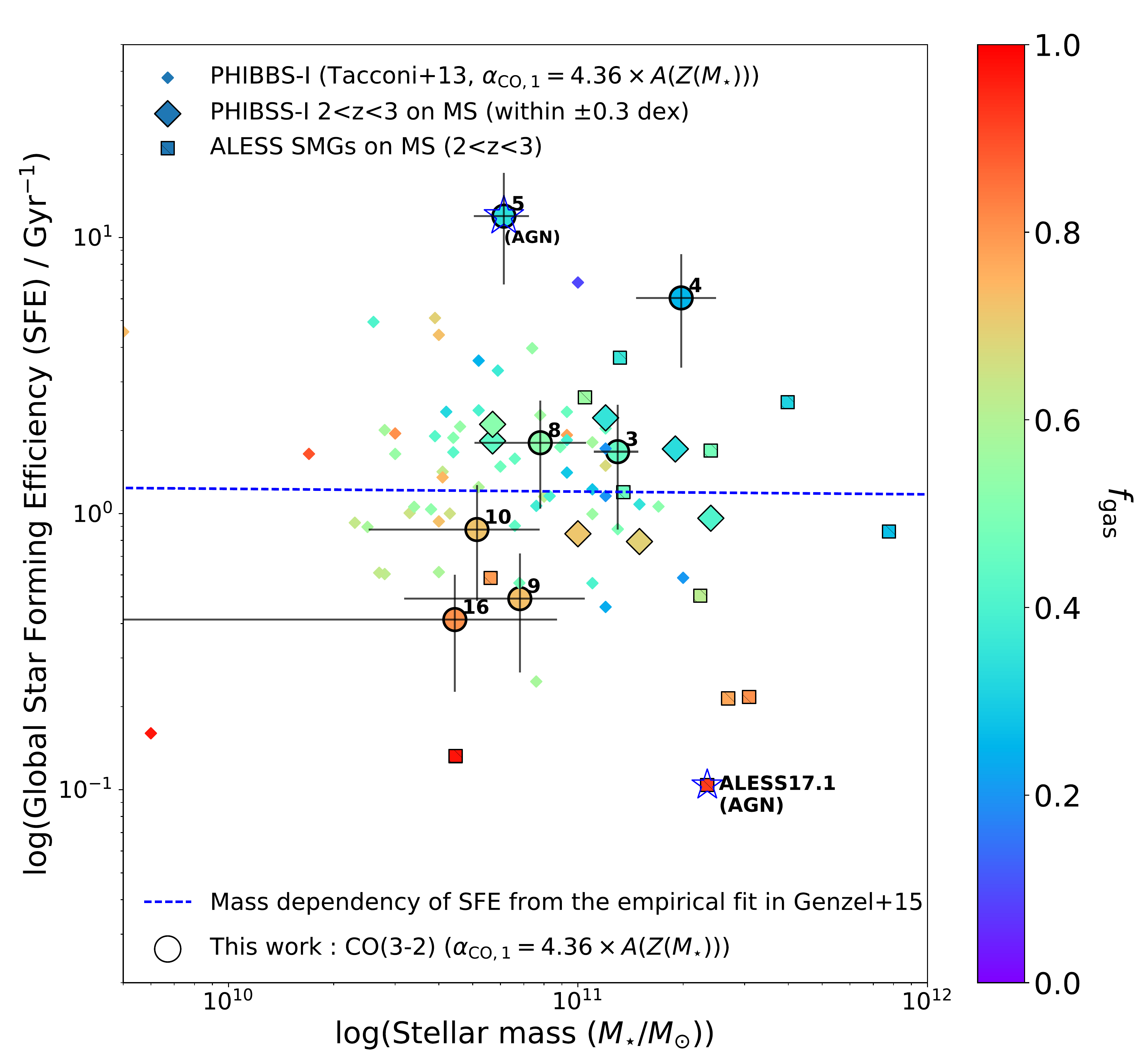}
\caption{Stellar-mass dependency of global star forming efficiency. Although the probed range is still narrow, a positive correlation between M$\star$ and SFE is found for the protocluster galaxies, in contrast to the result of \citealt{Genzel2015} for the main-sequence galaxies. 
The empirical fitting formula presented in \citealt{Genzel2015} shows a small dependency (with a power of 0.01) on stellar mass in the depletion time scale (i.e., a power of -0.01 for SFE indicated by the dashed line) compared to {the} larger contributions of the deviation from the main-sequence sSFR and the redshift evolution. 
The color scheme shows the $f_{\rm gas}$ for individual galaxies. 
A star symbol indicates the existence of AGN.
\label{fig:SFE-Mstar}
}
\end{figure*}
%%%%Figure end%%%%%%
%%%%%%%%%
%
\subsection{Environmental effect during the cluster-forming epoch?}
\subsubsection{Detection in the densest region}\label{sec:dense}
Motivated by the discovery of {the} morphology-density relation in \citet{Dressler1980}
and the relatively secure redshift ranges for the HAEs from the NB technique, 
we may regard the surface density as an indirect and rough representation of the cosmic web\footnote{If it were available, it would be better to discuss with 3D volume density. 
However, the number of galaxies is still too small to find a particular structure in 3D}.
The surface galaxy number density is measured by applying Gaussian kernels with a radius of $0^{\prime}.8$ ($\sim1.4$ comoving Mpc) and calculating the galaxy number within the area (Fig.~\ref{fig:fovs}).

Among seven CO-detected galaxies, five HAEs are located in the {region of} highest surface density.
We note that the number density of HAEs within the protocluster is three times larger than general fields (\citealt{Tanaka2011}, Tanaka et al., in preparation).
We plot the protocluster galaxies detected in Fig.~\ref{fig:fgas-sSFR} and colorize them to show the (relative) surface galaxy number densities, where the numbers next to the color bar are shown in the unit of arcmin$^{-2}$.
This suggests that the CO~(3--2) (or dust) detection of the galaxies depneds on (but not necessarily) the large scale structure.
\citet{Umehata2015, Umehata2017} argued that there is a concentration of the 1.1 mm continuum sources in the node of protocluster SSA22, where filamentary structures meet. 
If it also applies to our case, the detection in the region of highest surface density may be mirroring the preferential place of gas detection within certain large structures of the protocluster, e.g., projected filaments or the node, where the gas is infalling or being accreted. 
Future mapping observations of gas content
are necessary to visualize such phenomena, which will allow us to reveal environmentally driven galaxy evolution with differential gas supply (and consumption) at {a} high redshift.

If we focus on these five galaxies, they are again divided into two populations : (i) {a} vast amount of gas with {a} relatively low mass ($f_{\rm gas} \gtrsim 0.7$, $4\times10^{10}<M_{\star}/M_{\odot}\lesssim 1\times10^{11}$) in relatively less dense regions (HAE9, HAE10, and HAE16) and (ii) more massive galaxies ($\gtrsim 1\times10^{11}$) with lower $f_{\rm gas}\lesssim0.5$ in denser regions (HAE4 and HAE8).

\subsubsection{Comparison with previous studies}
To discuss further, we compare {out results} with recent results presented in \citet{Genzel2015} and \citet{Scoville2016}, in addition to our analysis of the ALESS SMGs on the main sequence and PHIBBS-I.
\citet{Genzel2015} enlarged the sample size by including the results of not only the PHIBBS-I galaxies but also PHIBBS-II, IRAM-COLDGASS, and other surveys.
Recently, \citet{Tacconi2017} presented extended work.
From these extensive studies, a scaling relation of the depletion time ($\tau_{\rm depl.}$) is empirically derived. 
The gas depletion time can vary with the redshift ($z$), offset from the main sequence {$\Delta$(MS)} and dependency of stellar mass {($M_{\star}$)}.
We use the empirical fit of \citealt{Genzel2015} and plot the expected line on the plane of $f_{\rm gas}$ versus sSFR in Fig.~\ref{fig:fgas-sSFR} by considering the definition of $f_{\rm gas}$.
The gas fraction can be equated with sSFR and $\tau_{\rm depl.}$, i.e., $f_{\rm gas}=1/(1+(sSFR\times\tau_{\rm depl.})^{-1}$. 
By the definition, at {a} fixed sSFR, $f_{\rm gas}$ decreases with decreasing $\tau_{\rm depl.}$.
We also fill an area using the characteristic depletion time $\tau_{\rm depl.}$ = 200-700 Myr for high-z galaxies, as presented in \citet{Scoville2016}, to show how $f_{\rm gas}$ changes as a function of sSFR with this depletion time range. 

Our targets and control samples (PHIBBS-I, ALESS SMGs on the main sequence) are within a narrow range near the main sequence. 
Therefore, at {an} {\it almost} fixed sSFR, the scattered points (the protocluster members, PHIBBS-I and ALESS SMGs on the main sequence) around the results of \citet{Genzel2015} and \citet{Scoville2016} can be regarded to indicate the dependence on the global SFE or $\tau_{\rm depl.}$(Fig.~\ref{fig:SFE-Mstar}).

By excluding AGN (HAE5), our results give a positive correlation between $M_{\star}$ and SFE (Pearson's correlation coefficient (in log scale) = 0.89 with a p-value of 0.02), and such a relation perhaps decreases the $f_{\rm gas}$ of massive galaxies, in contrast to the empirical fitting in \citet{Genzel2015} and \citet{Tacconi2017}, and the PHIBBS-I galaxies. 
The latter cases show a weak negative (or flat) correlation between $M_{\star}$ and SFE.
Although the intrinsic SFR of non-AGN, dusty galaxies (e.g., HAE9 and 16) might be higher than SFR$_{\rm H_{\alpha}}$(corrected) (Sec.~\ref{sec:extinction}), a positive correlation nevertheless holds (but becomes rather weaker).
One may argue that such {an} apparently different correlation from the general field is only due to the sample bias and {is} still explained within the scatter of the PHIBBS-I sample.
We cannot reject this argument with {the} current data set. 
This {issue} can only be investigated through larger and deeper observations
by collecting statistically large numbers.
However, we note that the SFE dependency of the stellar mass (and thus gas fraction)
appears to have some connection with the galaxy number density (Fig.~\ref{fig:fgas-sSFR}) and, thereby, perhaps with the environment, 
as shown in the previous section.

\subsubsection{Suggested picture and future aspects}
The correlations shown in the previous sections suggest some insight into massive galaxy evolution and the properties of dark matter within a high-z protocluster.

As discussed in \citet{Genzel2015} (section 4.3), 
the global depletion time can be related to dark-matter properties in the framework of disk formation within a dark-matter-dominated universe (\citealt{Mo1998} see also eq.(24) in G15) : the baryons' angular-momentum parameter ($\lambda$), galaxy's (local) star-formation efficiency ($\eta$), dark matter concentration parameter ($C_h$), and Hubble parameter ($H(z)$).
The similarity of the averaged physical properties for the galaxies on the main sequence may be due to two dominant factors between the balance of $H(z)$ and perhaps $\eta$.
The mass dependency of SFE (at an almost constant sSFR) suggests that additional (or different) physical processes, which are perhaps related to the environment, are necessary to explain this phnomenon.
Considering that halo concentration is higher in denser environments and increases in later times (\citealt{Bullock2001}) and that a (proto)cluster is a place where galaxy evolution proceeds earlier (e.g., having a quenched or passive population at the center in advance ; \citealt{Kurk2009, Strazzullo2013, Koyama2014, Cooke2016}), the halo concentration parameter of the most massive galaxies in denser regions might be higher than that of less massive galaxies in less dense regions which should be tested with future observations.

However, we note that the CO line widths tend to decrease with increasing stellar mass (and thereby $f_{\rm gas}$ (Table~\ref{tab:physicparam} or see the spectra in Fig.~\ref{fig:postage1}-\ref{fig:postage4}) and SFE), hinting at a change of the angular momentum parameter, which needs to be investigated with future higher-resolution observations with a constraint on the inclination.
Considering the total gas content is almost constant, as shown in Fig.~\ref{fig:fgasMstar} for the detected sources, this further shows a signature of changes in the dark matter fraction.
A test on the time scale of these changing parameters, is additionally required through both observations and simulations.
A quantitative estimation of all of these contributions may not be simple, but it is certainly required in future observations for understanding the environmental effect in galaxy evolution at high $z$.\\

The proposed picture, however, may be different for galaxies with stellar mass less than $10^{10}\, M_{\odot}$, provided non-detection, and they have to be investigated via deeper observation.
Thus far, there is little evidence for the change of the scatter of {the} main sequence in {a} different environment 
 (e.g., \citealt{Peng2010, Koyama2013, Darvish2016}), but recently, \citet{Hayashi2016} reported a larger (upward) scatter of main-sequence galaxies in the low-mass galaxies ($<10^{9.3} \; M_{\odot}$) at {the} $z=2.5$ protocluster, which is in fact similarly seen in our sample (I. Tanaka et al., in preparation; see Fig.~\ref{fig:galaxysequence}). 
%%%%%%%
\section{Conclusion}\label{sec:summary}

In this paper, we investigated the gas content of HAEs that are typical star-forming galaxies on the main sequence associated to protocluster \objectname{4C23.56} at z = 2.49. 
To derive the gas properties, we conducted CO~(3--2) (Band~3) and 1.1 mm ($\lambda_{\rm rest} \sim 385 \mu$m) dust continuum (Band~6) observations with ALMA toward the protoclusters for which panchromatic studies are available.
This is the first paper in a series of papers that reveal the gas properties of galaxies within the protocluster.
From the ALMA observations, our results are as follows.
\begin{enumerate}
\item We obtained seven CO~(3--2) and four 1.1 mm dust continuum {detections}. 
All four 1.1 mm detections are included in CO~(3--2){ detections}. 
While the parent galaxies have {a} stellar mass range greater than three orders of magnitude ($\log(M_{\star}/M_{\odot}) = [8, 11.5]$), the detected sources are all massive ($M_{\star}>4\times 10^{10}$ $M_{\odot}$) on the star-forming main sequence.
\item Gas mass was derived using the ``Galactic" conversion factor with additional correction for the metallicity dependence of the CO conversion factor using {the method described in} \citealt{Wuyts2014} and following the analysis presented in \citealt{Genzel2015} for CO~(3--2), which yields a consistent value derived from dust-based calibration using \citet{Scoville2016}.
\item The HAEs having either CO~(3--2) and 1.1 mm detection carry, on average, a gas mass content similar to those of main-sequence galaxies in general fields. 
The massive HAEs ($M_{\star}>4\times 10^{10}$ $M_{\odot}$) have {a} gas content in the range of (0.3-1.8) $\times 10^{11}$ M$_{\odot}$ and a median gas fraction $\langle f_{\rm gas} \rangle = 0.53 \pm 0.07$ for CO~(3--2) and 0.50 $\pm$ 0.06 for 1.1 mm measurement. 
Including our work, {the} high-$z$ massive galaxies ($2<z<3$) on the main sequence that were considered (\citealt{Tacconi2013, Scoville2016, Hodge2013, da Cunha2015}) all possess {a} higher gas content than {those of} local star-forming galaxies, regardless of their environment.

\item The cosmic gas density of high-$z$ protoclusters was measured for the first time. 
Using either the redshift range of CO~(3--2) ($\Delta z\sim0.01$) or NB filter ($\Delta z \sim$ 0.03), which is comparable with the predicted size in simulations (e.g., \citealt{Chiang2013}), and the survey area of Band~3, we found an enhancement of cosmic gas density, $\rho_{H_2}\sim (1-5) \times 10^{9} \,M_{\odot}\, {\rm Mpc}^{-3}$ that is already a factor of 6-22 higher value only with the detection than the upper limit set by the recent survey toward HUDF (\citealt{Decarli2016a}) with the same assumption of conversion factor and line ratio.

\item We found that $f_{\rm gas}$ decreases with increasing stellar mass, as observed in control samples. 
However, our sample differs in that $f_{\rm gas}$ also changes with surface galaxy number density.
Galaxies with {a} higher gas fraction ($f_{\rm gas}>0.7$) are less massive ($4\times10^{10}<M_{\star}/M_{\odot}\lesssim 1\times10^{11}$) in {regions with} relatively low surface density, while galaxies with $f_{\rm gas}\lesssim0.5$ are more massive ($\gtrsim 1\times10^{11}$) and in {regions with} higher surface density.

\item Massive main-sequence galaxies in the protocluster may be evolving under the effect of the specific environment. A systematically different correlation between SFE versus stellar mass might be the combined result of {a} higher gas volume density and {the} non-negligible contribution of dark matter imprinted in the surface number density (and CO line widths), but quantitative assessment should be performed in future studies {to confirm this hypothesis}.

\end{enumerate}

The sample size is still small to discuss statistical significance as a general picture of galaxy evolution. And the different methods used in the derivation of {parameters} other than M$_{\rm gas}$, i.e., SFR and $M_{\star}$, when comparing field samples.
Therefore, larger surveys are necessary to probe a wide range of characteristic environments (e.g., diverse galaxy number densities) and redshifts
that can be constructed with the same analysis tools.
Deeper observations are also necessary to investigate the evolution of less massive galaxies and {their} connection to the probed massive galaxies on the protocluster.

\acknowledgments
This paper makes use of the following ALMA data: ADS/JAO.ALMA\#2012.1.00242.S. ALMA is a partnership of ESO (representing its member states), NSF (USA) and NINS (Japan), together with NRC (Canada) and NSC and ASIAA (Taiwan) and KASI (Republic of Korea), in cooperation with the Republic of Chile. The Joint ALMA Observatory is operated by ESO, AUI/NRAO and NAOJ.
M. L. and T.S. were financially supported by a Research Fellowship from the Japan Society for the Promotion of Science for Young Scientists.
R.K. and Y.T. were supported by {a} JSPS Grant-in-Aid for Scientific Research (A) Number 15H02073.
{K.K. was supported by JSPS Grant-in-Aid for Scientific Research (A) Number 25247019.} 
T.K. is financially supported by Grant-in-Aid for Scientific Research (JP21340045 and JP24244015) from the Japanese Ministry of Education, Culture, Sports and Science.
Data analysis was in part carried out on the open use data analysis computer system at the Astronomy Data Center, ADC, of the National Astronomical Observatory of Japan.

\facilities{ALMA, ASTE (AzTEC), Spitzer (MIPS), Subaru (MOIRCS)}.
\software{
CASA \citep{casa2007} , 
Astropy \citep{astropy2013}
          }

\appendix
\renewcommand\thefigure{\thesection\arabic{figure}}  
\section{Flux versus S/N}\label{app:flux}
We also checked whether the peak flux is consistent with other flux measurements, i.e., Gaussian fitting (using CASA commend \texttt{imfit}) and aperture photometry that is clipped below the $2.6~\sigma$. 
For {the} CO~(3--2) measurement, we also tested the spectra-based fitting by integrating the spectrum by using CASA \texttt{specfit}, if it is available for each smoothed image. 
All {data} are measured from the primary beam corrected maps. 
Furthermore, except {for} HAE16 in Band~6, all of the sources are within a good sensitivity region.
While some compact sources (in the original image) with {a} high S/N ($\sim$ 10) have peak flux values {consistent} with {those of} other methods, {the} relatively low S/N (S/N$\leq 7 $) with extended ALMA detections do not ensure {that} a Gaussian fitting is a secure method. 
Therefore, taking a peak flux would be a more robust method to maximize the S/N and consider a galaxy as an unresolved source. 

We investigate the growth curve of a galaxy to optimize the smoothing kernel and then to estimate a flux (Fig.~\ref{fig:growthcurveb3} and ~\ref{fig:growthcurveb6}). 
The growth curve gradually approaches the maximum value, while the S/N reaches a peak and then decreases as {the} noise level increases and the smoothing Gaussian kernel becomes larger. 
In some cases, the peak flux decreases after it has reached a peak because of contamination in side-lobes in interferometric data sets or contamination from nearby galaxies (on the map). 
Smoothing major axis = 0.0 implies no smoothing. 
We tested growth curves using {kernels of} 0$^{\prime\prime}$.4 to 4$^{\prime\prime}$.0 in {steps of} 0$^{\prime\prime}$.2 (convolved beam size = 0.8 to 4.1) for Band~6 and 0.6 to 4.0 for Band~3 (convolved beam size = $\sim$ 0$^{\prime\prime}$.9 to 4$^{\prime\prime}$.1). 
Combining all {the} growth curves in Band~3 and Band~6, we decided to use the Gaussian kernel of $0^{\prime\prime}.8$ for Band~6 and $0^{\prime\prime}.6$ for Band~3.
%%%%
\setcounter{figure}{0}
\begin{figure}[htb!]
\centering
\includegraphics[width = 0.7\textwidth, bb = 0 0 1024 798]{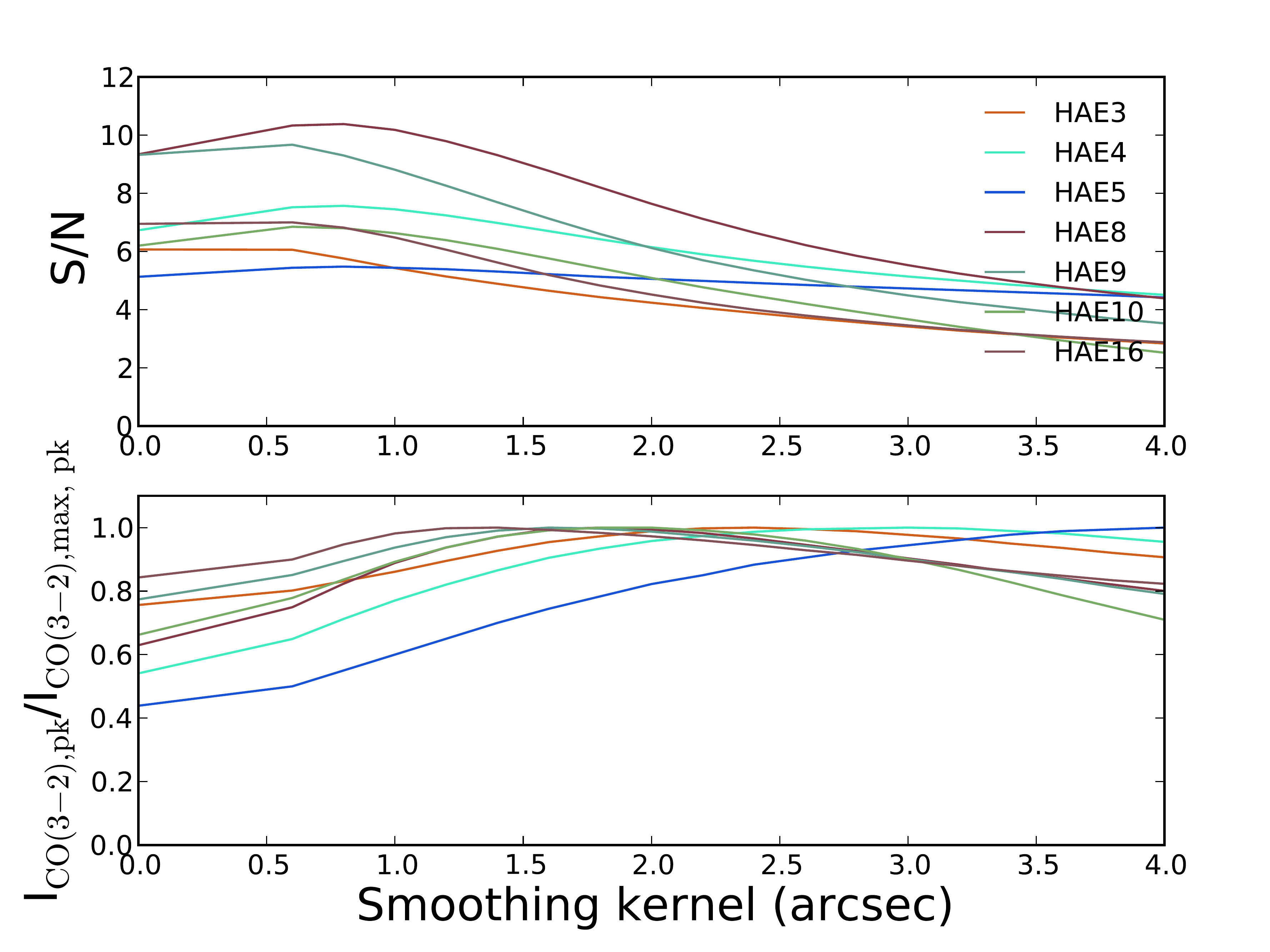}
\caption{S/N and peak flux growth curve in Band~3. We need to consider both S/N and flux to optimize the smoothing parameter to estimate a total flux. We have chosen a kernel of 0.6 $^{\prime\prime}$ to conduct uniform analysis with Band~6 data as well. At this kernel, the expected flux recovered at least 50\% of the maximum flux (but with low S/N).}
\label{fig:growthcurveb3}
\end{figure}

\begin{figure}[htb!]
\centering
\includegraphics[width = 0.7\textwidth, bb = 0 0 1024 798]{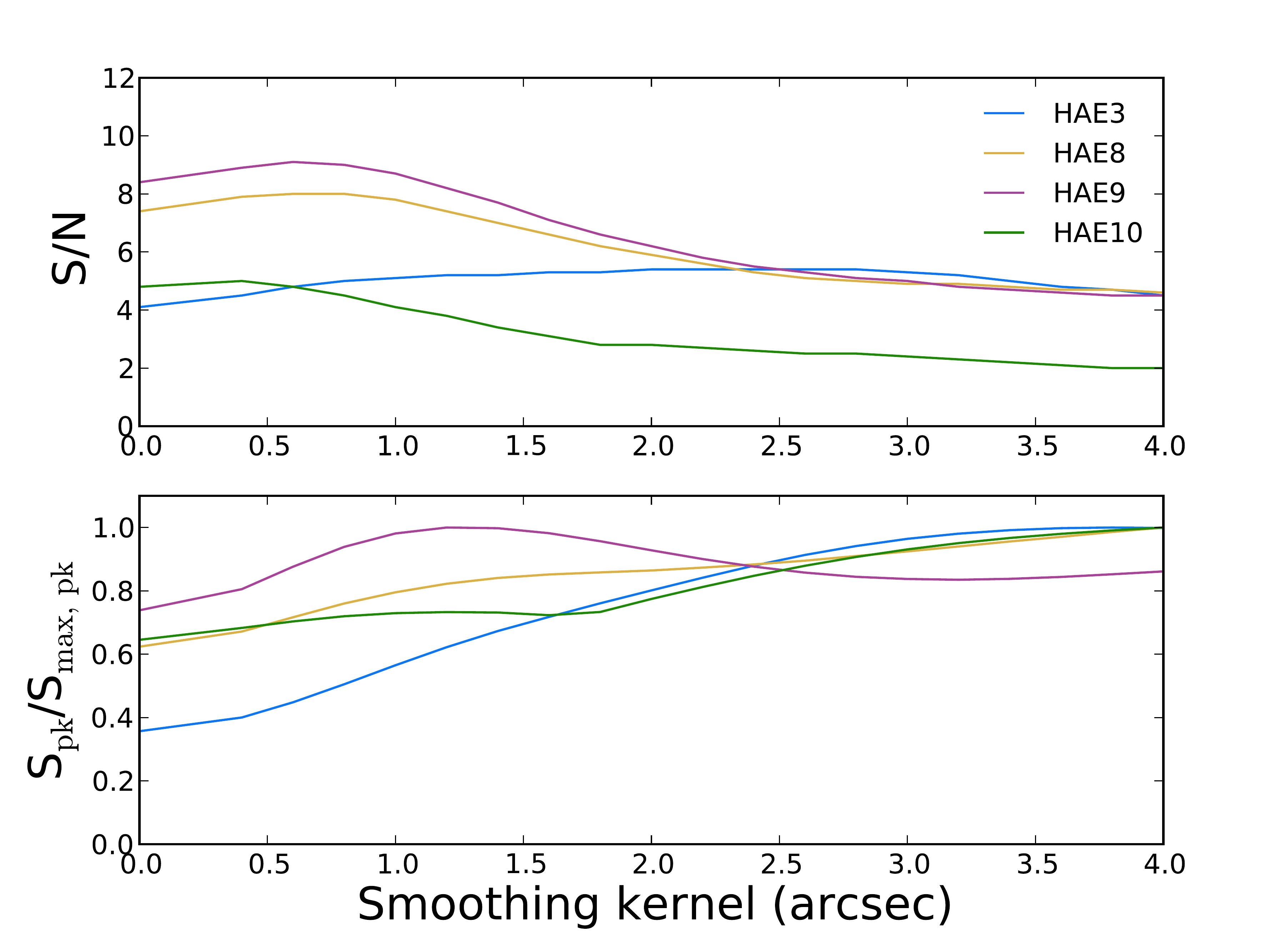}
\caption{Same growth curve {as in Fig.~\ref{fig:growthcurveb3}} but for Band~6. Again, although the peak flux is recovered {less} with the adopted kernel of $0.8^{\prime\prime}$, they have {a} low S/N, suggesting large uncertainties are also clearly included in the brightest peak.}
\label{fig:growthcurveb6}
\end{figure}
\section{Position error}{\label{app:positionerr}}
We investigated the peak position consistency between the H$\alpha$ position (I. Tanaka et al., in preparation) and CO~(3--2) or 1.1 mm. 
The observations have {a} similar resolution of $\sim$ 0$^{\prime\prime}$.7-0$^{\prime\prime}$.9. 
Figure~\ref{fig:positionerror} shows how far the peak position is offset in CO~(3--2) and 1.1 mm images with respect to the NB position. 
The position error expected from the interferometric data depends on {the} S/N and synthesized beam size. 
The expected position error is $\sim$ 0$^{\prime\prime}$.1. 
More errors that could be associated with the phase error in the phase calibrator {may be added}. 
Compared to this, the position accuracy for NB compared to 2MASS is $0^{\prime\prime}.044$ (I.Tanaka et al., in preparation).

While we conclude that the position is {\it roughly} consistent with each other within $\sim0^{\prime\prime}.4$ resolution, we note that there might be a systematic offset in the peak position of CO(3-2) and 1.1 mm compared to {the} H$\alpha$ peak (on average $\sim0^{\prime\prime}.2$). 
The source {with the highest offset} is HAE4 (see also Fig.~\ref{fig:postage1} ; the distribution of H$\alpha$ is extended compared to the distribution of CO(3-2) or 1.1 mm). 
The position difference between the H$\alpha$, CO~(3--2) and 1.1 mm, therefore, appears to originate from the difference in the internal structure of a galaxy and/or the effect of dust extinction. 
Further discussion should be conducted with {a} higher resolution and high sensitivity observation.

\setcounter{figure}{0}    
\begin{figure}[htb!]
\centering
\includegraphics[width = 0.7\textwidth, bb = 0 0 800 800]{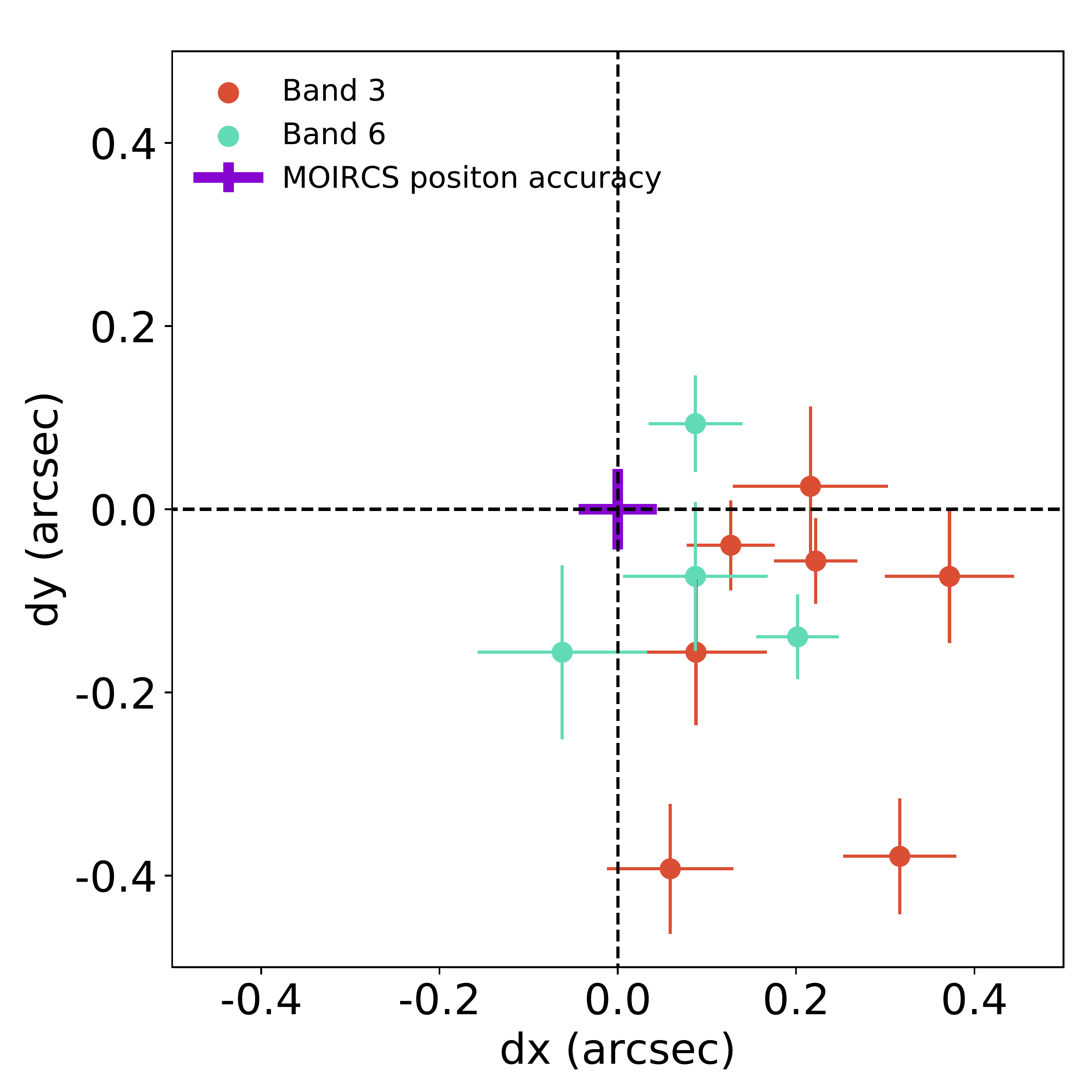}
\caption{Position offset with respect to {the} NB catalogue. We find that the position is roughly consistent with each other within $\sim0^{\prime\prime}.4$. Torquious circles indicate Band~6 1.1 mm {observations} and crimson circles indicate Band~3 CO~(3--2) {observations}.The position accuracy for NB compared to 2MASS is $0^{\prime\prime}.044$ (I. Tanaka et al., in preparation).}
\label{fig:positionerror}
\end{figure}

\section{Dynamical mass problem}\label{app:dynamical}
While being a fairly crude estimation, we compare a dynamical mass with the sum of the stellar {mass}, gas {and} and dark matter (DM) {mass}, to impose a limit on the conversion factor.
We estimate the dynamical mass by taking an average of two different estimators (i.e., an isotropic virial estimator and a rotating disk estimator) (\citealt{Tacconi2008, Tan2014}). 
 Both estimators scale the dynamical mass as a linear function of the galaxy size.
We adopt {a size larger by} a factor of two ($r_e$ = 10 kpc) than {compared to} the typical CO(3-2) size of a star-forming galaxy at $z\sim 2$ ($r_e$ = 5 kpc; e.g., \citealt{Tacconi2013, Bolatto2015}).
The exact size measurement will be presented in a subsequent paper, but the assumption adopted here is {meant} to provide one of the representative cases in our sample that are on the main sequence but require a lower conversion factor. 
In other words, the assumption provides upper limits of the dynamical mass, thereby placing upper limits on the conversion factor.
Without quantitatively addressing the size measurements, there are several supportive aspects for the assumption.
First, most sources are unresolved in CO~(3--2), H$\alpha$, and stellar component, with {a resolution of} $0^{\prime\prime}.7-0^{\prime\prime}.9$. 
Therefore, most of them presumably {have} (as of visual inspection) $r_e\lesssim10$ kpc.
Second, it is known that galaxy sizes measured from rest-frame UV and optical {spectra} (which trace a star-forming region and stellar component, respectively) decrease with increasing redshift (e.g., \citealt{Trujillo2007, van der Wel2014}) ; therefore, we do not expect extremely large ($r_e>10$ kpc) massive galaxies at such high redshift even if we take into account the size-mass relation. 
%Furthermore, if the UV emission is heavily obscured, particularly for those in the central region, the size estimation gives larger size than the true size.
Third, the CO measurements in \citet{Tacconi2008} and \citet{Bolatto2015} suggest that the gas distribution in both CO(3-2) and CO(1-0) is comparable to those observed from rest-frame UV and optical {bands}. 
Our visual inspection also supports such {a} picture (i.e., it rules out {the argument} that the sizes are significantly different from each other), but it could be the case that the CO(3-2) size {is larger} than {the} compact ($r_e\lesssim1$ kpc) stellar component.
Finally, although there are some arguments regarding the effect of the environment on the size difference particularly for high-$z$ clusters, in the case of quiescent early-type galaxies (e.g., \citealt{Rettura2010, Raichoor2012, Delaye2014}) and perhaps larger sizes in star-forming galaxies compared to those in general fields selected as Lyman-break galaxies (M. Kubo et al., in private communication), the derived galaxy sizes are still not extreme cases with $r_e>10$ kpc.
For the DM mass estimation included within $r_e$, we adopt a nominal value of 0.25 $M_{\rm dyn}$ (\citealt{Daddi2010})\footnote{See also \citet{Wuyts2016} for {a} slightly higher value or \citet{Price2016} for {a} smaller value ; all the values are in the range of $M_{\rm DM} \simeq (0.1-0.3) \times M_{\rm dyn}$, and we note that recent two studies have estimated the gas mass without direct measurement and instead with scaling relations.
In any case, the dark matter is less likely the dominant component within the effective radius.}. 
We used the line widths listed in Table~\ref{tab:physicparam}, which are likely the upper limits of FWHM that yield the highest S/N when integrating across the velocity range (following \citealt{Seko2016} ; M. Lee et al., in preparation). 
Since we cannot constrain the inclination of the galaxy from the measurement, substantial uncertainties {are} included in the estimation of true line width. 
Nevertheless, more than half of the line widths already exceed 400 km s$^{-1}$ ; therefore, we may be observing these galaxies edge-on, rather than face-on, for the disk structure. 
Otherwise, they must be extremely unstable.
We found the median dynamical mass $M_{\rm dyn}\sim 5 \times 10^{11}$ M$_{\odot}$,  and the masses of two out of seven HAEs (HAE5 and HAE8) are within the effective radius ($=(M_{\star} + M_{\rm gas})/2$){, which} already exceeds the estimated dynamical mass without considering {the} DM contribution, if we adopt the `Galactic' conversion factor $\alpha_{\rm CO} = 4.36$ (corrected for helium) or {an} even higher value of $\alpha_{\rm CO} = 6.5$ as suggested in \citet{Scoville2016}. 
The two galaxies have the lowest line widths. Our measurements may underestimate the dynamical mass within the effective radius.
If we adopt a loose constraint on the size, $r_e=6$ kpc, HAE4 {would also be a} galaxy that cannot be explained by the dynamical estimator.
Although the uncertainty included in the measurements {is} large, we caution that some galaxies might tend to reduce the conversion factor below the assumed value.
%

%\LongTables
%\begin{landscape}
\clearpage
\startlongtable
\movetabledown=1.6in
\begin{rotatetable}
\begin{deluxetable}{cccccccccccccc}
\tabletypesize{\scriptsize}
\tablecaption{Source information for detection\label{tab:physicparam}}
\tablewidth{0pt}
\tablehead{
\colhead{Source ID} & \colhead{RA$_{\rm CO32}$} & \colhead{Dec$_{CO32}$ } & 
\colhead{$M_{\star}$} & \colhead{SFR$_{H\alpha, corr}$} & \colhead{$z_{\rm CO~(3-2)}$} & \colhead{I$_{\rm CO~(3-2)}$} & line width &
\colhead{S$_{\rm 1.1 mm}$} & \colhead{M$_{\rm gas,\, CO32}$} & \colhead{M$_{\rm gas,\, dust}$} & \colhead{SFE} & \colhead{f$_{\rm  gas,\,CO}$} &  \colhead{ID in T11}
\\
\colhead{} & \colhead{ (J2000)} & \colhead{(J2000)} & 
\colhead{$\times$ 10$^{10}$ M$_{\odot}$} & \colhead{M$_{\odot}$ yr$^{-1}$} & \colhead{} & \colhead{Jy km s$^{-1}$} & \colhead{km s$^{-1}$}&
\colhead{mJy} & \colhead{$\times$ 10$^{10}$ M$_{\odot}$} & \colhead{$\times$ 10$^{10}$ M$_{\odot}$} & \colhead{Gyr$^{-1}$} &\colhead{}
 & }
\startdata
HAE3 & 316.837650 & 23.520500 & 13.0$\pm$1.9 & 176$\pm$78 & 2.4861 & 0.352$\pm$0.06 & 500 & 0.53$\pm$0.14 & 10.55$\pm$1.8 & 5.9$\pm$2.16 & 1.68   &   0.45  &  354\\
HAE4 & 316.840213 & 23.527986 & 19.7$\pm$5.1 & 414$\pm$175 & 2.478 & 0.246$\pm$0.03 & 300 & $<$0.54 & 6.86$\pm$0.84 & $<$6.22 & 6.04   &   0.26  &  479\\
HAE5 & 316.820742 & 23.508458 & 6.1$\pm$1.1 & 374$\pm$140 & 2.4873 & 0.09$\pm$0.02 & 100 & $<$0.33 & 3.14$\pm$0.7 & $<$3.96 & 11.95   &   0.34  &  153\\
HAE8 & 316.816433 & 23.524292 & 7.8$\pm$2.7 & 156$\pm$63 & 2.4861 & 0.263$\pm$0.03 & 300 & 0.75$\pm$0.12 & 8.69$\pm$0.99 & 8.35$\pm$2.51 & 1.81   &   0.53  &  431\\
HAE9 & 316.844121 & 23.528694 & 6.8$\pm$3.6 & 90$\pm$40 & 2.4861 & 0.542$\pm$0.06 & 1000 & 1.21$\pm$0.21 & 18.43$\pm$2.04 & 13.48$\pm$4.14 & 0.49   &   0.73  &  511\\
HAE10 & 316.815525 & 23.520000 & 5.1$\pm$2.6 & 115$\pm$47 & 2.4861 & 0.362$\pm$0.06 & 500 & 0.44$\pm$0.12 & 13.15$\pm$2.18 & 4.9$\pm$1.83 & 0.88   &   0.72  &  356\\
HAE16 & 316.811025 & 23.520958 & 4.4$\pm$4.3 & 76$\pm$32 & 2.4826 & 0.493$\pm$0.07 & 600 &$ <$1.1 & 18.52$\pm$2.63 & $<$12.5 & 0.41   &   0.81  &  --\\
\enddata
\end{deluxetable}
\end{rotatetable} 
%%%%%
%%%%%%
\begin{deluxetable}{cccccccccccccc}
\tabletypesize{\small}
\tablecaption{Information for undetected sources\label{tab:nondet}}
\tablewidth{0pt}
\tablehead{
\colhead{Source ID} & \colhead{RA$_{H\alpha}$} & \colhead{Dec$_{H\alpha}$ } & 
\colhead{S$_{\rm CO32}$}\tablenotemark{a}  & \colhead{S$_{\rm 1.1 mm}$} & \colhead{M$_{\rm gas, \,CO32}$} & \colhead{M$_{\rm gas,\, dust}$} &  \colhead{ID in T11}\\
\colhead{} & \colhead{ (J2000)} & \colhead{(J2000)} & 
\colhead{mJy} & \colhead{mJy}&
\colhead{$\times$ 10$^{10}$ M$_{\odot}$} & \colhead{$\times$ 10$^{10}$ M$_{\odot}$}
}
\startdata
HAE1 & 316.811658 & 23.529211 & $<0.67$   &  $<0.32$   &  $<11.07$   &  $<3.54$ & 491\\
HAE2 & 316.840738 & 23.530434 & $<0.52$   &  $<0.50$   &  $<8.61$   &  $<5.60$ & 526\\
HAE6 & 316.839548 & 23.522090 & $<0.95$   &  $<0.46$   &  $<15.66$   &  $<5.08$ & 393\\
HAE7 & 316.814680 & 23.527065 & $<0.52$   &  $<0.86$   &  $<8.52$   &  $<9.56$ & --\\
HAE12 & 316.812222 & 23.529876 & $<0.67$   &  $<0.32$   &  $<11.12$   &  $<3.56$ & --\\
HAE13 & 316.840917 & 23.528263 & $<0.49$   &  $<0.43$   &  $<8.02$   &  $<4.79$ & 500\\
HAE14 & 316.832414 & 23.514173 & $<0.53$   &  $<0.70$   &  $<8.68$   &  $<7.79$ & --\\
HAE15 & 316.833151 & 23.518959 & $<0.52$   &  $-$\tablenotemark{b}   &  $<8.56$   &  $-$\tablenotemark{b} & --\\
HAE17 & 316.823395 & 23.530683 & $<1.02$   &  $<1.07$   &  $<16.92$   &  $<11.93$ & 543\\
HAE18 & 316.840110 & 23.533663 & $<0.7$   &  $-$\tablenotemark{b}    &  $<11.54$   &  $-$\tablenotemark{b}  & --\\
HAE19 & 316.842465 & 23.529443 & $<0.5$   &  $<0.37$   &  $<8.21$   &  $<4.12$ & --\\
HAE20 & 316.812277 & 23.522381 & $<0.65$   &  $<1.01$   &  $<10.67$   &  $<11.20$ & --\\
HAE21 & 316.811748 & 23.528571 & $<0.63$   &  $<0.33$   &  $<10.45$   &  $<3.67$ & --\\
HAE22 & 316.824409 & 23.529090 & $<1.03$   &  $-$\tablenotemark{b}    &  $<17.09$   &  $-$\tablenotemark{b} & --\\
HAE23 & 316.811469 & 23.521843 & $<0.71$   &  $<1.10$   &  $<11.71$   &  $<12.27$ & -- \\
\enddata
\tablenotetext{a}{At 100 km s$^{-1}$ resolution per channel. 3$\sigma$ upper limit}

\tablenotetext{b}{ALMA 1.1 mm observation has no coverage}

\end{deluxetable}

\end{document}